\RequirePackage{fix-cm}
\documentclass[smallextended]{svjour3}       
\smartqed  
\usepackage{graphicx}
\usepackage{textcomp}
\usepackage{gensymb}
\usepackage{aas_macros}
\usepackage[intlimits]{amsmath}
\usepackage{hyperref}

\newcommand*{\addthinspace}{\hskip0.16667em\relax}
\begin{document}

\title{A compact instrument for gamma-ray burst detection on a CubeSat platform I
}
\subtitle{Design drivers and expected performance}

\author{David Murphy \and Alexey Ulyanov \and Sheila McBreen \and Maeve Doyle \and Rachel Dunwoody \and Joseph Mangan \and Joseph Thompson \and Brian Shortt \and Antonio Martin-Carrillo \and Lorraine Hanlon
}


\institute{D. Murphy 
            \and
            A. Ulyanov \and S. McBreen \and M. Doyle \and R. Dunwoody \and J. Mangan \and A. Martin-Carrillo \and L. Hanlon
            \at
            School of Physics \& Centre for Space Research, University College Dublin, Dublin 4, Ireland\\
            \email{david.murphy@ucd.ie}
            \and
            J. Thompson
            \at
            School of Mechanical and Materials Engineering \& Centre for Space Research, University College Dublin, Dublin 4, Ireland
            \and
            B. Shortt
            \at
            European Space Agency, ESTEC, 2200 AG Noordwijk, The Netherlands
}

\date{Received: date / Accepted: date}

\maketitle

\begin{abstract}
The Educational Irish Research Satellite 1 (EIRSAT-1) is a 2U CubeSat being developed under ESA's Fly Your Satellite! programme.
The project has many aspects, which are primarily educational, but also include space qualification of new detector technologies for gamma-ray astronomy and the detection of gamma-ray bursts (GRBs).
The Gamma-ray Module (GMOD), the main mission payload, is a small gamma-ray spectrometer comprising a 25\,mm\,$\times$\,25\,mm\,$\times$\,40\,mm cerium bromide scintillator coupled to an array of 16 silicon photomultipliers.
The readout is provided by IDE3380 (SIPHRA), a low-power and radiation tolerant readout ASIC.
GMOD will detect gamma-rays and measure their energies in a range from tens of keV to a few MeV.

Monte Carlo simulations were performed using the Medium Energy Gamma-ray Astronomy Library to evaluate GMOD's capability for the detection of GRBs in low Earth orbit.
The simulations used a detailed mass model of the full spacecraft derived from a very high-fidelity 3D CAD model.
The sky-average effective area of GMOD on board EIRSAT-1 was found to be 10\,cm$^2$ at 120\,keV.
The instrument is expected to detect between 11 and 14 GRBs, at a significance greater than 10$\sigma$ (and up to 32 at 5$\sigma$), during a nominal one-year mission.
The shape of the scintillator in GMOD results in omni-directional sensitivity which allows for a nearly all-sky field of view.


\end{abstract}

\section{Introduction}
\label{intro}

Gamma-ray bursts (GRBs) are intense flashes of gamma radiation which originate from distant galaxies and typically last from a fraction of a second to several minutes \cite{vedrenne2009}.
Thousands of GRBs have been detected to date by high-energy astronomy missions such as the Compton Gamma Ray Observatory\cite{goldstein2013}, High Energy Transient Explorer \cite{hete}, Neil Gehrels Swift Telescope \cite{Gehrels_2004}, Fermi Space Telescope \cite{Meeg2009ApJ...702..791M,Atwood_2009} and INTEGRAL \cite{Wink2003A&A...411L...1W}.
The distribution of GRB durations demonstrates a bi-modality, with two classes of GRBs associated with different generation mechanisms \cite{kouveliotou1993}.
Long GRBs (\textgreater2\,s) are produced in the core-collapse of massive stars \cite{macfayden1998} while short GRBs (\textless2\,s) are associated with mergers of compact binary systems \cite{1984SvAL...10..177B}.

The detection of the short gamma-ray burst GRB 170817A~\cite{goldstein2017} in coincidence with the gravitational wave (GW) signal GW170817 from a binary neutron star inspiral~\cite{abbott2017a} marked the beginning of a new era of multi-messenger astronomy and experimentally confirmed that at least some of the progenitor systems for short GRBs are binary neutron star mergers.
The joint localisation of the GRB and GW event led to multi-wavelength follow-up observations of the GRB afterglow, providing information on the orientation of the binary system, and the detection of the associated kilonova in ultraviolet, optical and infrared band~\cite{followup_Abbott_2017}. 
To date, 67 GW candidates have been recorded by the LIGO~\cite{2015CQGra..32g4001L} and Virgo~\cite{2015CQGra..32b4001A} interferometers, including several neutron star--neutron star and neutron star--black hole merger candidates~\cite{2020arXiv201014527A,gws_nature}.
However, no electromagnetic counterparts have been detected for any of these events except the aforementioned GW170817.

The GW170817/GRB 170817A discovery and its follow-up campaign emphasised the importance of simultaneous gamma-ray and GW observations.
This demand will be further increased by future major upgrades of the GW observatories \cite{2018LRR....21....3A} which will improve their sensitivity and the detection rate of GW events.
The detection of $10_{-10}^{+52}$ binary neutron star mergers, of $1^{+91}_{-1}$ neutron star--black hole mergers and $79^{+89}_{-44}$ binary black hole mergers in one calendar year is predicted~\cite{gw_prospects_2020} for the next operating run (O4) of LIGO~\cite{2015CQGra..32g4001L}, Virgo~\cite{2015CQGra..32b4001A} and KAGRA~\cite{2019NatAs...3...35K}, which is planned to commence in July 2022.

However, a major challenge to the detection of electromagnetic counterparts is the potential lack of future gamma-ray missions.
Many of the current missions, including the Neil Gehrels Swift Telescope, Fermi Space Telescope and INTEGRAL, are all approaching or have exceeded their nominal mission lifetimes. Two major GRB-related missions are in a study  phase, THESEUS~\cite{amati2018} and AMEGO~\cite{mcenery2019}.

Current and planned near-future large scale missions, including SVOM \cite{svom_2015arXiv151203323C}, a Chinese-French mission due for launch in 2022, will not provide the full-sky coverage required for efficient detection of electromagnetic counterparts to GW events \cite{perkins2017}.
This potential gap has led to a search for alternative solutions, such as a fleet of small satellites with gamma-ray detecting technology.
A Chinese mission, GECAM \cite{zhang2019}, utilises two GRB detecting small satellites to provide full-sky coverage.
Each satellite uses a hemispherical array of LaBr$_3$ and SiPM instruments, achieving GRB localisation capability of a few degrees.
The pair launched in December 2020 and detected its first GRB in January 2021 \cite{2021GCN.29331....1A}.
Other agencies are in the process of developing CubeSats with GRB detection and localisation capabilities e.g. BurstCube \cite{perkins2017}, MoonBEAM \cite{moon_hui_briggs_2018}, HERMES \cite{fiore2020,evangelista2020,sanna2020,campana2020}, CAMELOT \cite{werner2018} and GRID \cite{wen2019}.
GRBAlpha \cite{pal2020} is an in-orbit demonstration for CAMELOT which has recently been launched \cite{grbalphalaunch}.
To detect gamma-rays, all of the above mentioned CubeSat missions use inorganic scintillator crystals (CsI:Tl, NaI:Tl, GAGG:Ce) read out by silicon photomultipliers (or silicon drift detectors in the case of HERMES).
The miniature size and low weight of these photosensors allow the CubeSats to carry relatively large detectors with an effective area on the order of 100\,cm$^2$, comparable to the effective area of the Fermi GBM detectors \cite{Meeg2009ApJ...702..791M}.
CubeSats are relatively low cost and have short launch timescales \cite{OnthevergeofanastronomyCubeSatrevolution} making them ideal candidates to bridge potential gaps in coverage by the large gamma-ray missions.
In addition, a fleet of GRB detecting CubeSats could independently provide detection and localisation of GRBs during future gravitational wave instrumentation operating runs \cite{fuschino2019}. 

CubeSats are a class of nano-satellite (usually defined as having a mass of less than 10\,kg) which approximately conform to the Cal Poly CubeSat specification \cite{cubesat_spec_rev13}.
They may be a number of different, approximately cuboid, sizes which are all built from predefined configurations of multiples of the 1U CubeSat unit.
The base CubeSat unit (the 1U CubeSat) has at its core a 10\,cm $\times$ 10\,cm $\times$ 10\,cm cubic volume.
The Cal Poly specification allocates a mass of 1.33\,kg per 1U volume for most CubeSat sizes, though the new 6U specification allows for a total mass of 12\,kg. 

The Educational Irish Research Satellite 1 (EIRSAT-1) is a 2U CubeSat which will be Ireland's first satellite~\cite{ssea18_eirsat}.
EIRSAT-1 was proposed in response to an ESA announcement of opportunity as part of their educational Fly Your Satellite! (FYS!) programme and was accepted into that programme in 2017.
The FYS! programme supports university student teams to build, launch, and operate their own CubeSat and has launched more than 11 CubeSats since 2008.
It is an initiative of the ESA Academy which provides educational opportunities for university students \cite{ssea18_eirsat,Doyle2020,Dunwoody2020,Walsh2020,ssea18_adm,ssea18_wbc}.
The main goal of the proposal was to fly and space-qualify a novel gamma-ray detector that was already under development~\cite{ulyanov2016,ulyanov2017}.
This detector had a technologically mature design using a cerium bromide (CeBr$_3$) scintillator and silicon photomultipliers (SiPMs), and had demonstrated compatibility with a CubeSat form factor.
The detector was capable of measuring gamma-rays in 30\,keV--10\,MeV energy range which would make it suitable for GRB detection.
However, the detector would require a compact low-power readout system, such as the SIPHRA ASIC, to fit in a CubeSat and therefore the EIRSAT-1 Gamma-ray Module (GMOD) was envisaged as a demonstrator to combine all the necessary components in a CubeSat payload. 
Such a payload is seen as a significant advance over legacy instrumentation which typically relied on classical photomultiplier tube scintillator detection and discrete control and readout electronics.
A prototype configuration of the payload was evaluated on a balloon flight \cite{murphy2021} and used to test the radiation harness of the SiPMs \cite{ulyanov2020}.

This paper gives an overview of the GMOD instrument to be flown on board EIRSAT-1 and presents its expected in-orbit performance obtained from Monte Carlo simulations.
The detailed detector design and experimental characterisation of the module are described in a separate paper~\cite{gmod2}.
The final detector design has been environmentally qualified \cite{Mangan2021}.

%
\section{EIRSAT-1}
\label{sec:eirsat}

EIRSAT-1 is a 2U CubeSat measuring 22.7\,cm\addthinspace$\times$\,10\,cm\addthinspace$\times$\,10\,cm, incorporating GMOD as its primary payload along with two other novel payloads --- EMOD (discussed below) and Wave-Based Control \cite{ssea18_wbc,thompson2016}, a novel control algorithm which will be used in an experimental attitude control capacity.
To ensure mission success, design decisions were made to prioritise mission reliability and lifetime over a complex mission profile.
As part of the FYS! programme, EIRSAT-1 is planned to be deployed from the International Space Station (ISS), resulting in an expected mission lifetime of approximately one year (9--18\,months depending on Solar activity).

\begin{figure}
\includegraphics[width=\textwidth]{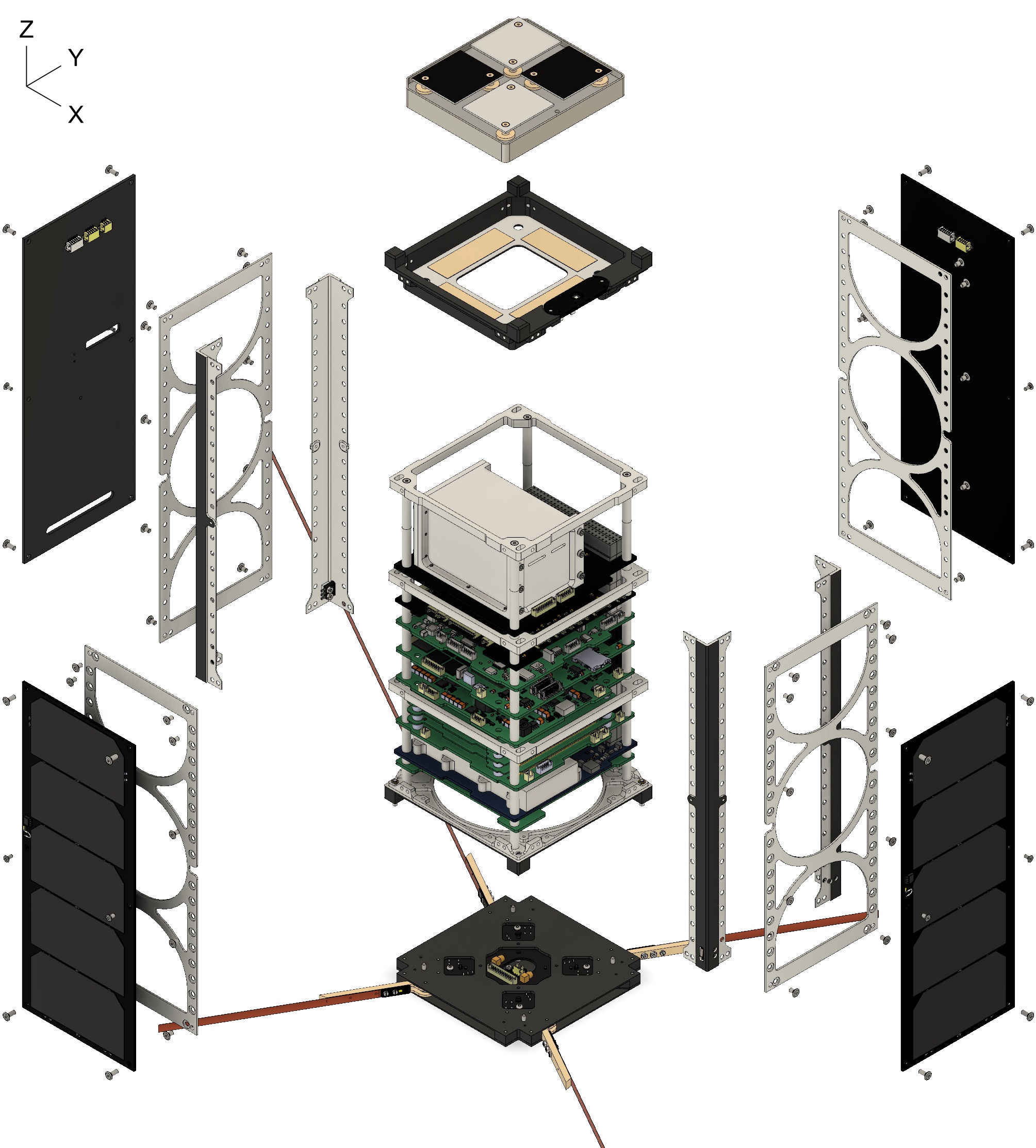}
\caption{Exploded isometric view of the EIRSAT-1 spacecraft. GMOD, comprising a motherboard PCB and a large cuboid detector assembly, is the top subsystem in the main stack shown at the centre of the drawing. The EMOD experiment can be seen above the stack.}
\label{fig:eirsat-1exploded}
\end{figure}

The overall spacecraft design is presented in Figure~\ref{fig:eirsat-1exploded} and is based around the requirements of accommodating the three payloads on a CubeSat platform, launched into an ISS-like orbit at 400\,km altitude.
The spacecraft is built around a core stack of electronic subsystems manufactured by \r{A}AC Clyde Space.
Details of the particular subsystems in use can be found in \cite{ssea18_eirsat}.
This subsystem stack is supported by a 2U \r{A}AC Clyde Space structure which has been heavily modified to suit the needs of the EIRSAT-1 mission, particularly the +Z end-cap which has been entirely replaced by a custom structural element to accommodate the EMOD payload.

The structure is surrounded by body-mounted solar arrays on four sides.
Due to the anticipated launch of the spacecraft from the ISS, achieving the maximum possible mission lifetime requires minimising drag, meaning that the use of deployable solar arrays is not possible and power generation is therefore limited.
The limited power budget precludes the use of reaction wheels, limiting EIRSAT-1 to magnetic attitude actuators, which in turn requires that the communication system work with the spacecraft in any orientation.
The communication is therefore based on a VHF/UHF system utilising omni-directional deployable antennas \cite{ssea18_adm} which can be seen at the -Z end of the spacecraft illustrated `below' the stack in Figure~\ref{fig:eirsat-1exploded}.

The GMOD detector sits at the `top' of the main stack.
It is surrounded on four sides by the body-mounted solar arrays.
Between GMOD and the solar arrays are 1\,mm thick aluminium structural shear panels though these are skeletonised and are primarily a clear aperture having little effect on GMOD.

The Thermal Coupon Assembly (TCA) of the EMOD payload is located in the spacecraft `above' GMOD.
This payload is an experiment to perform measurements of two thermal management coatings known as SolarBlack \cite{doherty2016} and SolarWhite \cite{doherty2016a} which were developed in support of ESA's Solar Orbiter mission.
The TCA contains samples of the coatings which are measured using resistance temperature detectors and which must be placed on the exterior of the spacecraft where they will be exposed to solar radiation and must be thermally isolated from the spacecraft.
To ensure good isolation, the samples are mounted on a 1\,mm thick titanium baseplate via PEEK columns with a multi-layer insulation blanket included between the samples and the baseplate.
At energies below approximately 50\,keV, this construction is expected to provide more effective shielding than the body-mounted solar arrays leading to larger effective area for GMOD in the X and Y directions than in the Z direction.

%
\section{GMOD --- The Gamma-ray Module}
\label{sec:gmod}

\begin{figure}
\includegraphics[width=0.8\textwidth]{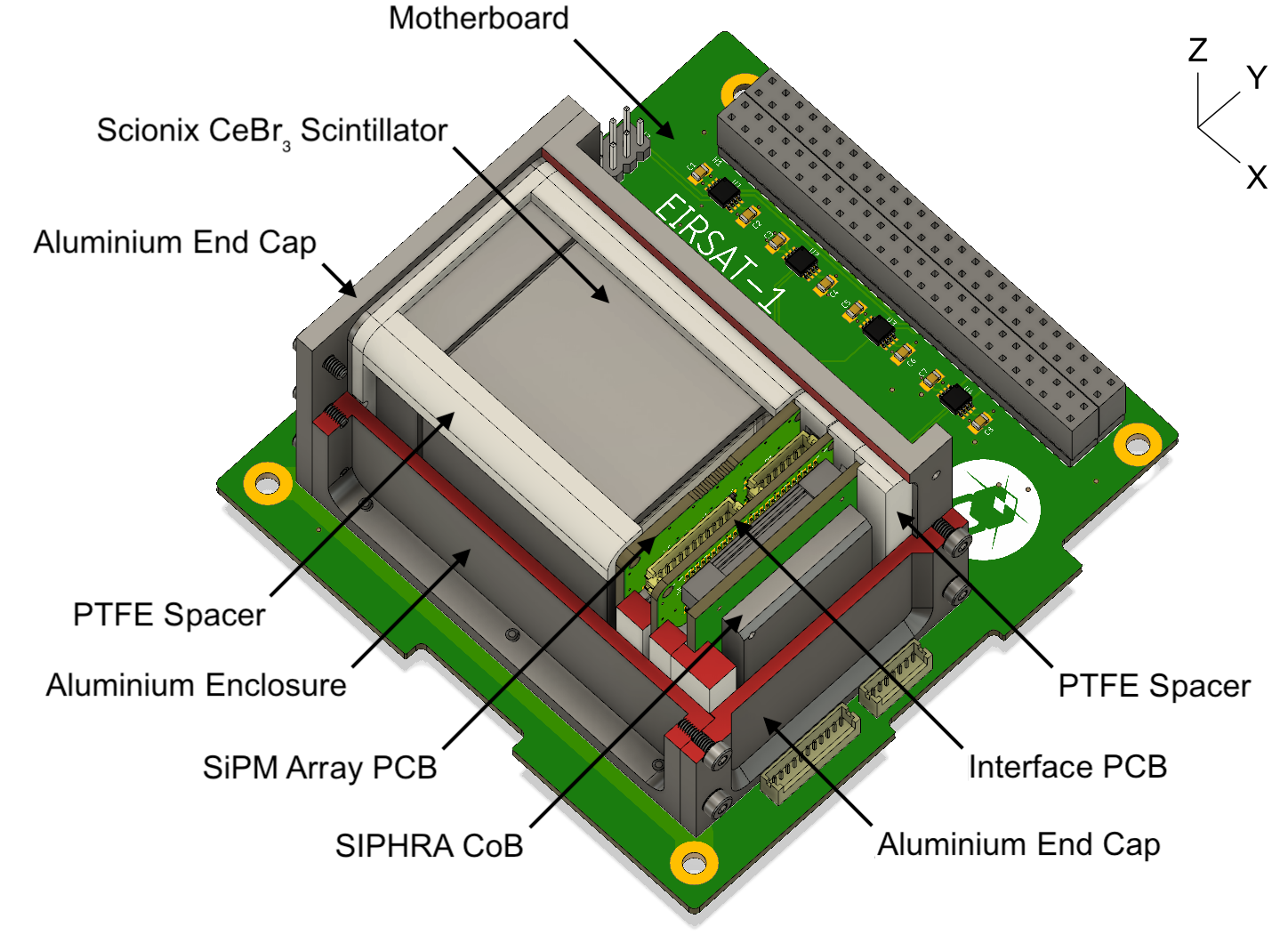}
\caption{Cut-away view of the GMOD payload. The design comprises a detector assembly mounted on a motherboard PCB. The detector assembly includes a CeBr$_3$ scintillator, SiPM array, and the SIPHRA ASIC within a light-tight enclosure. Red surfaces indicate a component which has been cut to reveal internal features.}
\label{fig:gmodchopped}
\end{figure}

GMOD is a scintillator-based gamma-ray detector which utilises a cerium bromide (CeBr$_3$) crystal scintillator, ON Semiconductor (formerly SensL) J-series silicon photomultipliers and the SIPHRA application specific integrated circuit (ASIC).
The instrument consists of a motherboard and a detector assembly which hosts the scintillator, SiPMs, and SIPHRA in a light-tight enclosure.
A cutaway view of GMOD illustrating its design can be seen in Figure~\ref{fig:gmodchopped} and a detailed description of the design and assembly can be found in \cite{gmod2}.

The scintillator is a 25\,mm\addthinspace$\times$\,25\,mm\addthinspace$\times$\,40\,mm CeBr$_3$ scintillator produced by Scionix.
As CeBr$_3$ is hygroscopic, the scintillator is supplied in an aluminium hermetically sealed unit with a quartz window.
The scintillation light from the crystal is measured by a custom 4$\times$4 array of 6\,mm J-series SiPMs \cite{jseries} which are readout and digitised by the SIPHRA ASIC \cite{ide3380} from Integrated Detector Electronics AS (IDEAS).
The SiPM array is assembled on one side of a dedicated `SiPM Array PCB'.
The reverse side of this PCB features supporting passive electronics for the SiPMs, the interface connectors for the board, and a PT100 temperature sensor which allows the array temperature to be monitored.
An `Interface PCB' sits between the SiPM array and SIPHRA and provides connectors for a harnessed connection to the motherboard.
PTFE spacers support the detector components within the aluminium light-tight enclosure, forming the 75\,mm\addthinspace$\times$\,51\,mm\addthinspace$\times$\,42\,mm detector assembly which is bolted to the motherboard.
These spacers additionally provide good thermal isolation between the enclosure and the internal components, stabilising the scintillator and SiPM array temperature \cite{Mangan2021}.

The motherboard includes the functionality necessary to interface the detector assembly to the spacecraft, including control and readout of the ASIC and regulation of the various voltages required by the detector including the SiPM bias voltage.
The SiPM bias PSU is adjustable between $-25$\,V and $-28.3$\,V allowing the SiPM over-voltage, and therefore the SiPM gain, to the adjusted in flight.
This voltage can also be automatically varied in response to the measured array temperature to maintain a constant gain, accounting for temperature-dependent variations of the SiPM breakdown voltage \cite{jseries}.
The motherboard has been designed to be compatible with the standard CubeSat PC-104 form factor and the whole GMOD assembly is placed at the top of the main spacecraft stack of subsystems.
The GMOD detector assembly and motherboard can be seen in the stack assembly in Figure~\ref{fig:eirsat-1exploded}.

SiPMs are known to receive damage from proton radiation leading to an increase in operating current and detector noise. These effects were evaluated using a prototype of the GMOD detector irradiated with a proton beam~\cite{ulyanov2020}. After one year of operation in the ISS-like orbit, the detector is still expected to detect gamma rays with energy above 30\,keV, which exceeds the mission requirement of 50\,keV. The total current of 16 SiPMs may increase to 500--1200S\,$\mu$A depending on temperature, which is easily handled by the power supply and SIPHRA readout. The SiPM radiation damage is not expected to be a problem for the GMOD detector as it uses a bright scintillator with a relatively small SiPM array and the SiPM exposure to protons in the ISS orbit will be relatively low (1\,MeV neutron equivalent fluence of $4.3\times 10^8$\,cm$^{-2}$ over the mission lifetime of one year). With a higher energy threshold of 50\,keV, the detector can achieve a one year operation in higher orbits.

\section{Simulations}
\label{ch:GMODsimulations}

To understand the performance of GMOD as a detector for gamma-ray bursts, its response was simulated using the Medium Energy Gamma-ray Astronomy Library (MEGAlib) \cite{zoglauer2006}. MEGAlib is a collection of software tools designed to simulate the performance of various gamma-ray instruments using the Geant4 simulation toolkit~\cite{AGOSTINELLI2003250} and to perform event reconstruction and analysis for Compton and pair production telescopes. This section describes the EIRSAT-1 mass model and the GRB and background models used in the MEGAlib simulations. The analysis of simulated data is discussed in Section~\ref{sec:simanalysis}.

%
\subsection{Mass Model}
\label{sec:massmodel}

The full EIRSAT-1 spacecraft has been included in the mass model as it is important to characterise the capabilities of the GMOD detector in the environment in which it will operate.
The mass model is derived from a very high-fidelity 3D CAD model of EIRSAT-1.
An exploded render of this model is shown in Figure~\ref{fig:eirsat-1exploded}.
The high-fidelity model is made up of contributions from various sources.
The models of the commercial off-the-shelf components were provided by \r{A}AC Clyde Space.
Models of the custom structural components were created in Autodesk AutoCAD.
The models of all GMOD mechanical components, including the scintillator crystal and hermetic enclosure (see Section~\ref{sec:gmod}) were also modelled in AutoCAD.
These models were also the basis for the manufacturing drawings that were sent to the workshop for fabrication.
To generate 3D models of EIRSAT-1's custom circuit boards, an EIRSAT-1 electronic components library was created in Autodesk EAGLE.
All components beyond extremely standard ones such as resistors and capacitors were managed using this library, primarily to ensure that all component footprints were correct based on manufacturer's drawings, but it also allowed 3D models of components to be added.
With all circuit components managed in EAGLE, it was possible to export high-fidelity 3D representations of the PCBs.
The 3D models of the Antenna Deployment Module and EMOD coupon assembly were provided by the respective sub-teams.

Due to the text-based mass model format required by MEGAlib, it is not possible to import 3D models from standard CAD formats and instead the volumes need to be defined manually.
Furthermore, only a limited number of volumes types are available in MEGAlib, including cuboid, sphere and cylinder (both of which may be solid or hollow and may be segmented), cone (which may also be hollow and be truncated), and various types of trapezoids.

\begin{figure}
\begin{center}
  \includegraphics[width=0.8\textwidth]{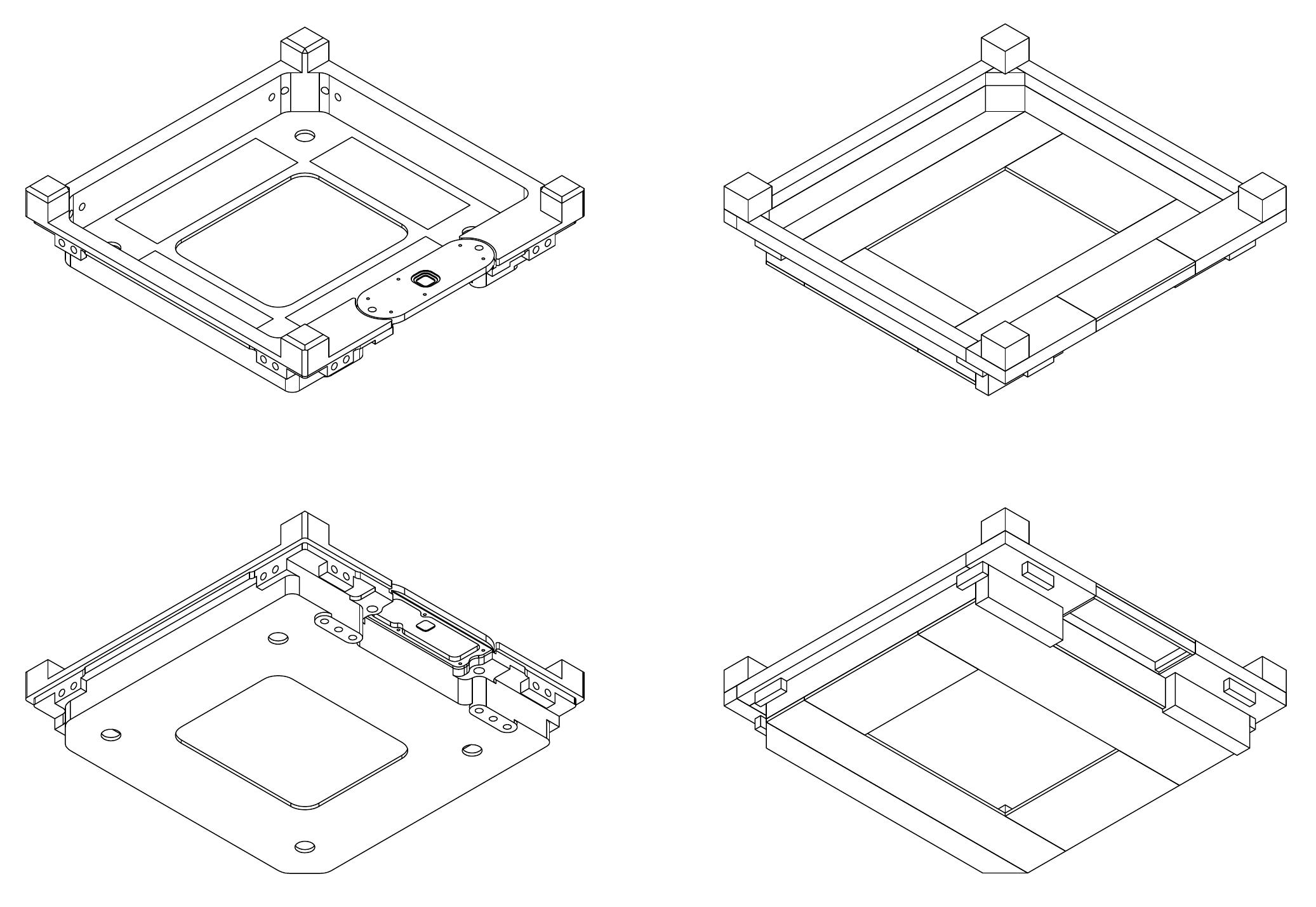}
  \caption[Simplified +Z End-cap for MEGAlib]{A comparison of the high-fidelity (left) and simplified for MEGAlib (right) models of the $+Z$ end-cap. The top row shows an isometric view of the top of the end-cap while the bottom row shows an isometric view of the underside of the end-cap.}
  \label{fig:simplifiedcap}
\end{center}
\end{figure}

\begin{figure}
\begin{center}
  \includegraphics[width=0.8\textwidth]{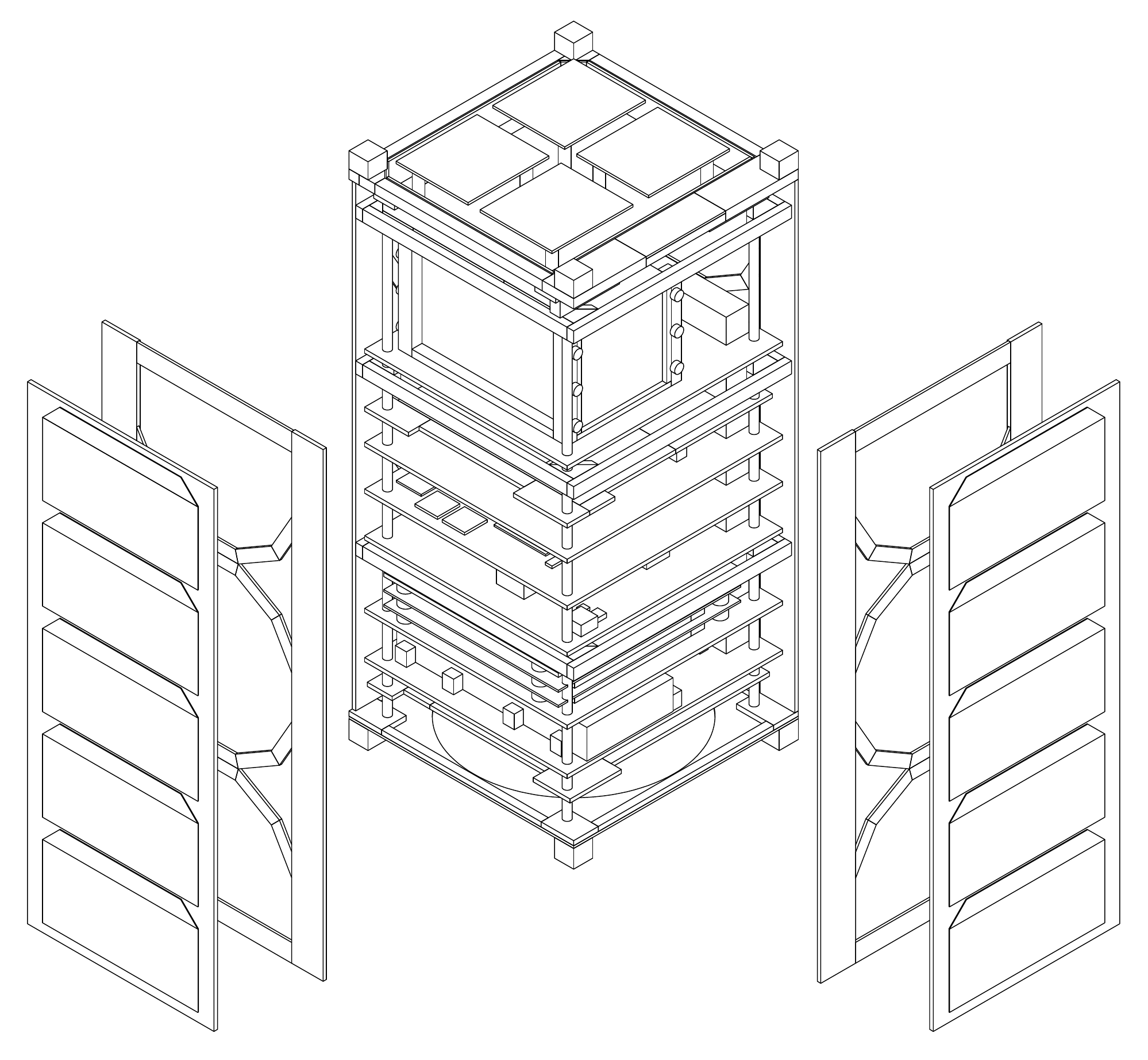}
  \caption[Simplified EIRSAT-1 for MEGAlib]{Partially exploded isometric view of the simplified AutoCAD 3D model of EIRSAT-1. The model has been designed such that all of the solids which comprise it can be represented as MEGAlib volumes.}
  \label{fig:simplifiedsat}
\end{center}
\end{figure}

The high-fidelity model of EIRSAT-1 was therefore simplified to the point that it contained only solids which could be represented using these supported MEGAlib volumes.
While this has the effect of removing finer details from the model, the overall mass distribution is not significantly changed and provides equivalent effective shielding. 
This simplification process was performed manually using AutoCAD, as it was significantly easier to place the volumes in a 3D environment and ensure that they correctly represented the EIRSAT-1 geometry using the AutoCAD interface than using the MEGAlib text-based volume description format.
For EIRSAT-1, the cuboid and trapezoid proved to be the most useful of the available volumes with many of the complex components of EIRSAT-1 being made of a combination of cuboids and trapezoids.
An example of this is given in Figure~\ref{fig:simplifiedcap}, which compares the $+$Z end-cap of the spacecraft before and after the simplification process and shows how the simplified version is constructed from multiple cuboids and trapezoids.
The full spacecraft simplified model is shown in Figure~\ref{fig:simplifiedsat}.

\begin{figure}
\centering
    \includegraphics[width=0.4\textwidth]{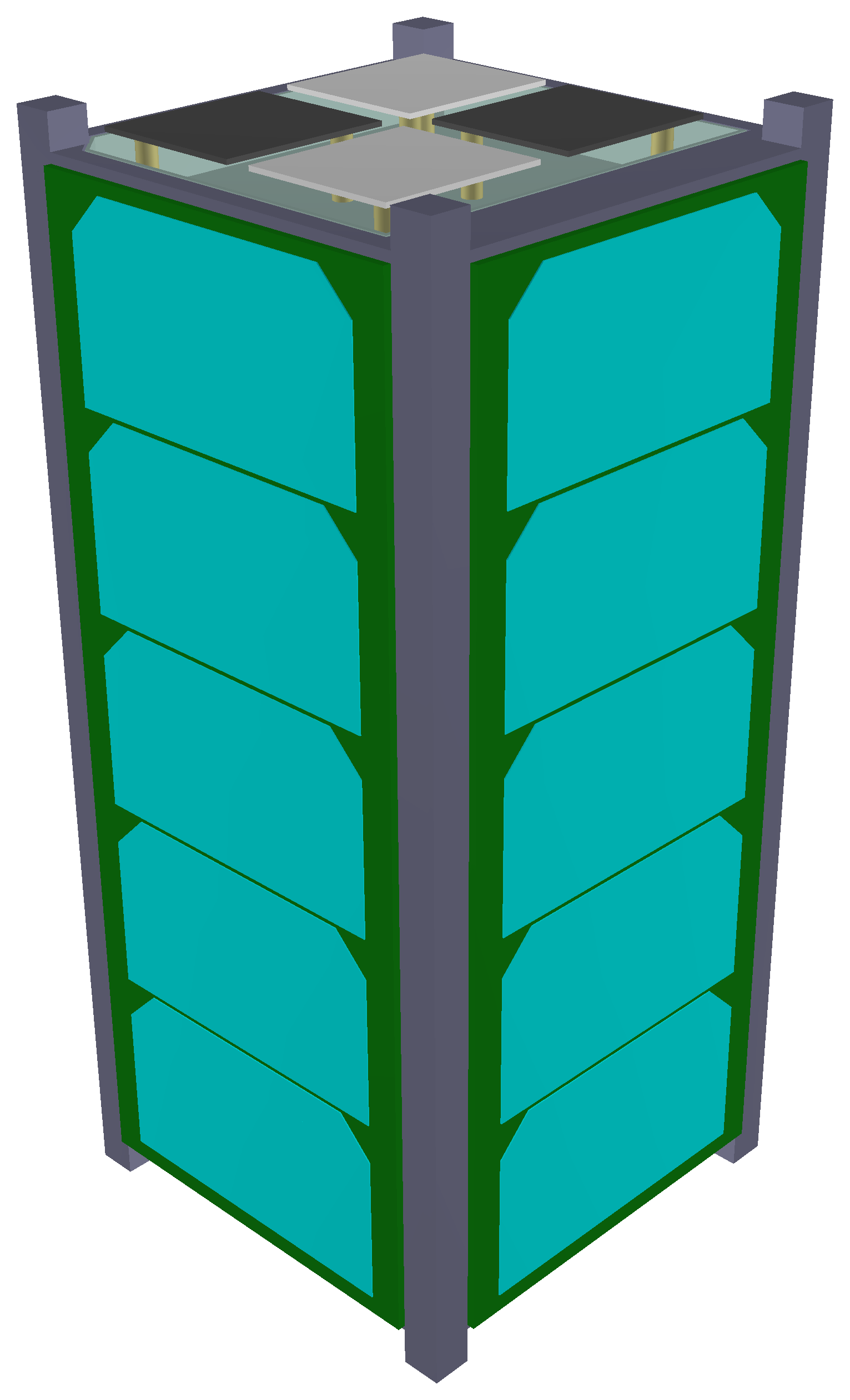}
    \hspace{1cm}
    \includegraphics[width=0.4\textwidth]{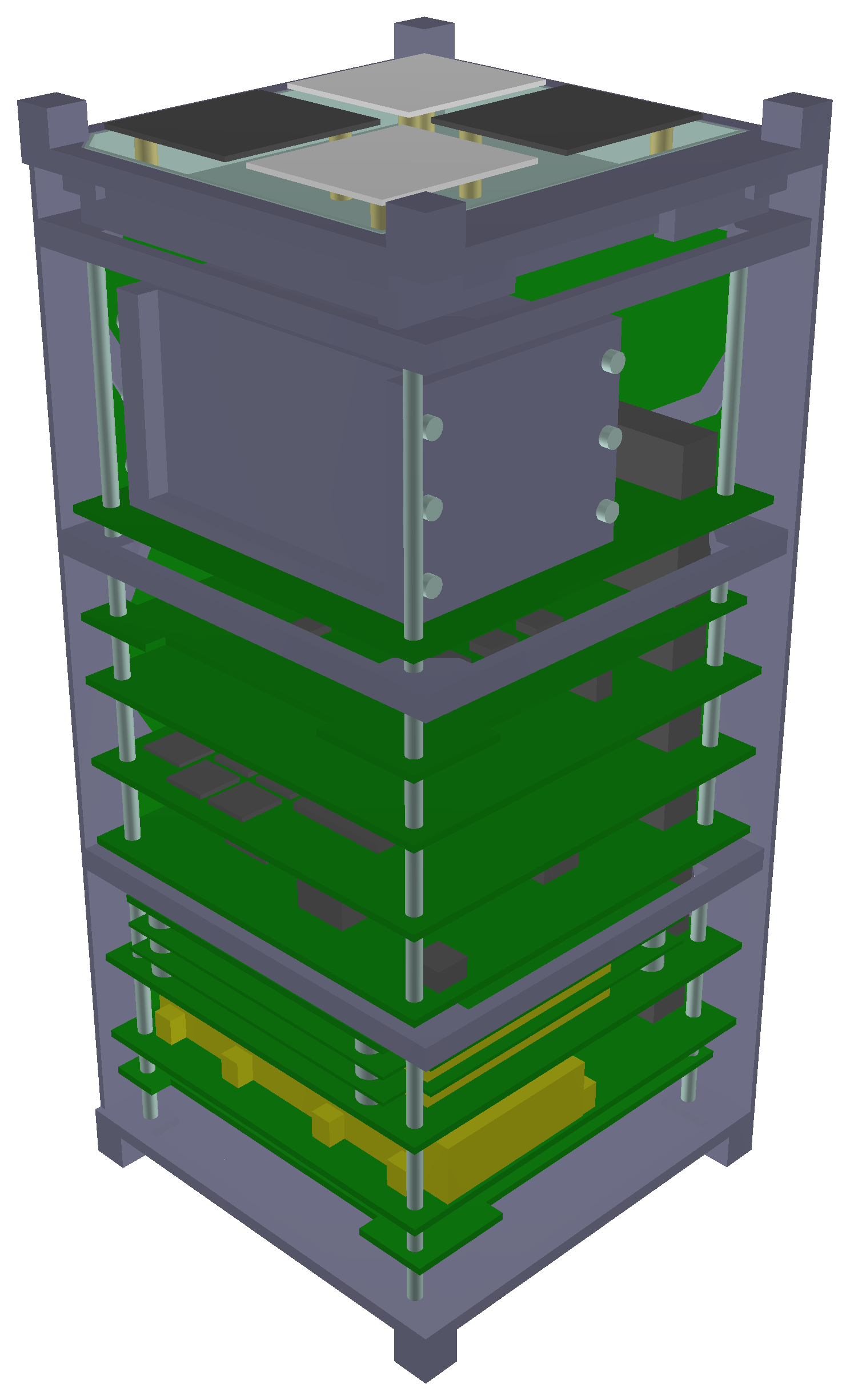}
    \par\bigskip\bigskip
    \includegraphics[width=0.4\textwidth]{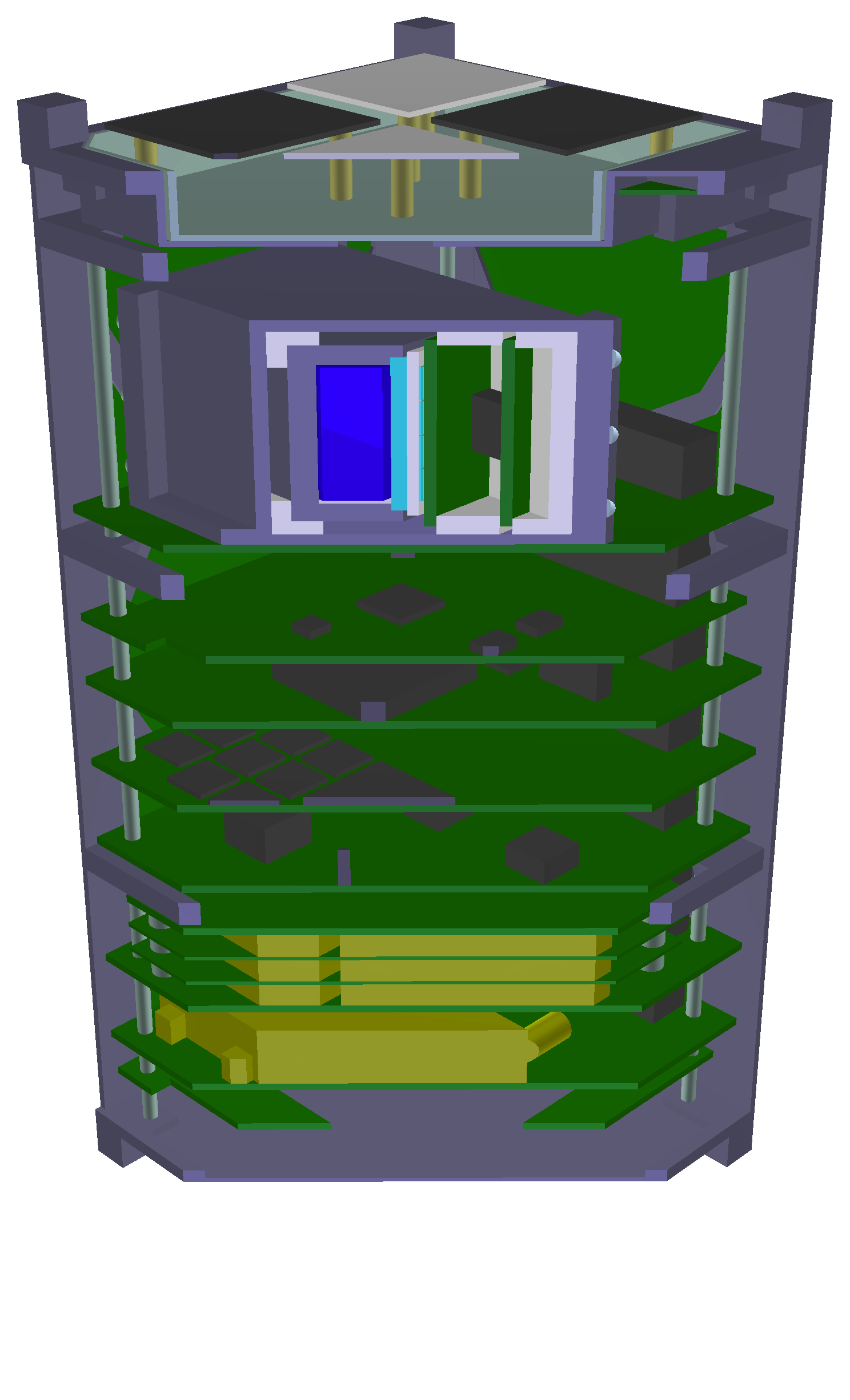}
    \hspace{1cm}
    \includegraphics[width=0.4\textwidth]{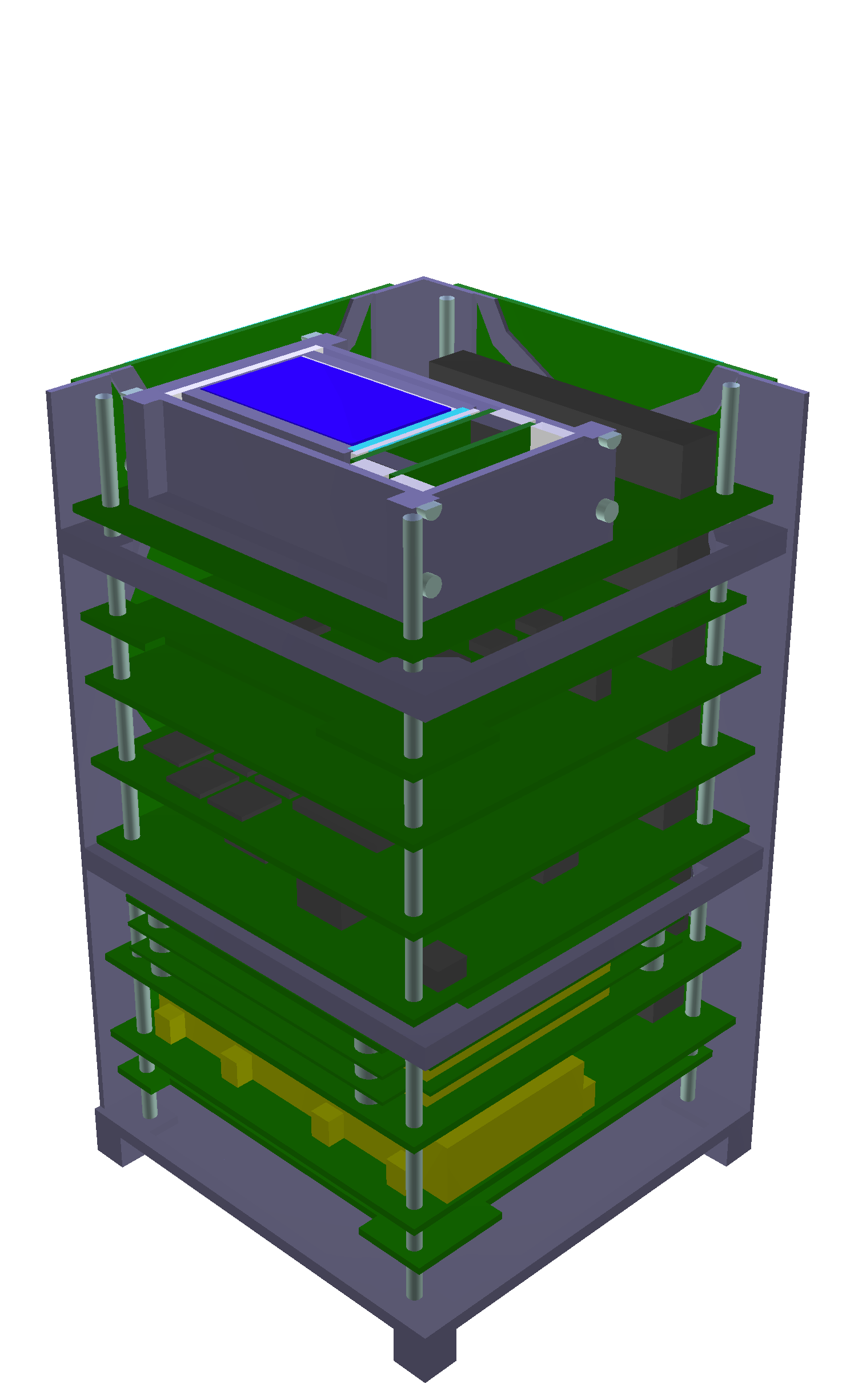}
  \caption[EIRSAT-1 MEGAlib Mass Model]{Various views of the EIRSAT-1 MEGAlib mass model showing the internal subsystems and constituent components of the GMOD detector assembly. \emph{Top-left:} Outside view of the full mass model. \emph{Top-right:} +X and $-$Y solar arrays and shear panels removed, showing internal components. \emph{Bottom-left:} Diagonal vertical slice through the spacecraft. \emph{Bottom-right:} Horizontal slice through the centre of the GMOD detector assembly.}
  \label{fig:massmodel}
\end{figure}

Following the simplification process, the spacecraft geometry was manually transcribed into the MEGAlib format by defining the position, orientation, size and material of each volume as well as relationships between volumes and mother volumes.
Renderings of the final mass model are shown in Figure~\ref{fig:massmodel}.
The mass model can be downloaded from \cite{eirsatmegalib}.

A GMOD MEGAlib detector was defined as the `Scintillator' type with a sensitive volume of 25\,mm\,$\times$\,25\,mm\,$\times$\,40\,mm which represented the CeBr$_3$ crystal.
The energy resolution of the detector was defined in the MEGAlib geometry file as Gaussian with the sigma values shown in Table~\ref{table:megalibresolution}.
These values were based on spectral measurements of gamma-rays from $^{137}$Cs and $^{22}$Na sources  using a development model of GMOD and are in good agreement with later measurements performed with the GMOD engineering qualification model \cite{gmod2}. The trigger threshold was set to 30\,keV.

\begin{table}[h]
\caption{GMOD energy resolution in the MEGAlib model}
\label{table:megalibresolution}

\begin{tabular}{l l}
\hline\noalign{\smallskip}
Energy (keV)  & 1$\sigma$ (keV)\\
\noalign{\smallskip}\hline\noalign{\smallskip}
20     &   8.5 \\
100    &   9.4 \\
350    &  12.1 \\
511    &  13.6 \\
662    &  14.9 \\
1000   &  17.6 \\
5000   &  41.5 \\
10000  &  67.0 \\
\noalign{\smallskip}\hline
\end{tabular}
\end{table}

\subsection{GRB Source Model}
\label{sec:grbmodel}

For the vast majority of GRBs, the spectral distribution of gamma-rays can be described by a smoothly broken power-law known as the Band function~\cite{band1993}:
\begin{equation}
  \label{eq:band}
  \frac{\mathrm{d}N}{\mathrm{d}E} =
      \begin{cases}
  A E^\alpha \exp\left(-\frac{(2+\alpha)E}{E_\mathrm{peak}}\right) & \text{if } E < \frac{\alpha-\beta}{2+\alpha}E_\mathrm{peak} \\
  A \left(\frac{\alpha-\beta}{2+\alpha}E_\mathrm{peak}\right)^{\alpha-\beta} \exp(\beta-\alpha) E^\beta & \text{if } E \geq \frac{\alpha-\beta}{2+\alpha}E_\mathrm{peak},
      \end{cases}
\end{equation}
where $A$ is the amplitude (normalisation factor of the spectrum),  $\alpha$ is the low energy index, $\beta$ is the high energy index and $E_\mathrm{peak}$ is the peak energy of the power density spectrum $E^2\mathrm{d}N/\mathrm{d}E$.

In this study, a Band function with parameter values $\alpha = -1.1$, $\beta = -2.3$ and $E_\mathrm{peak} = 300$\,keV was used to represent the spectrum of an average GRB, based on the BATSE spectral analysis of bright bursts~\cite{kaneko2006}.
The GRB spectrum was simulated within an energy range of 20\,keV--300\,MeV, as GMOD will only detect photons with measured energies above 20\,keV. 

To simulate the detector response to GRBs from all possible directions, the simulation was performed using an isotropic distribution of source gamma-rays over the full sky (using a source type `FarFieldAreaSource' in MEGAlib).
Directional point-like sources were then created in the analysis stage by filtering the simulation outputs and restricting the directions of the source gamma rays to small areas of the sky.

\subsection{Background Model}
\label{sec:bgmodel}

Operation of gamma-ray detectors in low Earth orbits is affected by many sources of background radiation, including cosmic gamma-rays, cosmic-ray particles and secondary particles generated in Earth's atmosphere. A recent summary of various background components and models can be found in \cite{cumani2019}. According to simulations~\cite{galgoczi2021}, the count rate of a scintillator detector in low Earth orbit outside the Van Allen radiation belts is dominated by the extragalactic gamma-ray background and atmospheric (albedo) gamma-ray emission. This is consistent with a background model for Fermi GBM based on real data~\cite{biltzinger2020}. The relative importance of cosmic and albedo gamma-rays depends on the photon energy band, with the cosmic gamma-ray background playing a major role at low energies and albedo photons dominating the gamma-ray spectrum above about 200\,keV, depending on the orbit altitude and inclination.

Only the two main background sources, that is cosmic and albedo gamma-rays, were considered in this work. The cosmic gamma-ray background was simulated using the model described in \cite{turler2010}, with the photon flux calculated as
\begin{equation}
  \label{eq:cgbmodel}
  F = \frac{0.109}{(E/28\,\mathrm{keV})^{1.40}+(E/28\,\mathrm{keV})^{2.88}}\ \mathrm{ph}\, \mathrm{cm}^{-2} \mathrm{s}^{-1} \mathrm{sr}^{-1} \mathrm{keV}^{-1}
\end{equation}

Albedo gamma-rays are generated through interactions of cosmic rays with the Earth atmosphere. Due to the effects of the geomagnetic field and solar activity on cosmic rays, the albedo flux seen by a satellite in a low Earth orbit can vary by a factor of 4--5 depending on the geomagnetic latitude and  solar activity~\cite{imhof1976,sazonov2007}. For this study we used the models given in \cite{sazonov2007} for $E<1$\,MeV and \cite{mizuno2004} for $E\geq1$\,MeV, with the photon flux in units of ph\,cm$^{-2}$\,s$^{-1}$\,sr$^{-1}$\,keV$^{-1}$ given by
\begin{equation}
  \label{eq:albedomodel}
  F =
      \begin{cases}
        \frac{C}{(E/44\,\mathrm{keV})^{-5}+(E/44\,\mathrm{keV})^{1.4}}\  & \text{if } E < 1\,\text{MeV} \\
        1.01\times 10^{-4} (\frac{E}{1000\,\mathrm{keV}})^{-1.34} & \text{if } 1\,\text{MeV} \leq E < 20\,\text{MeV} \\
        7.29\times 10^{-4} (\frac{E}{1000\,\mathrm{keV}})^{-2.0} & \text{if } E \geq 20\,\text{MeV},
      \end{cases}
\end{equation}
where the constant $C$ was set to 0.0080 to avoid discontinuity at 1\,MeV.
Equation~\ref{eq:albedomodel} gives the albedo photon flux for a geomagnetic latitude of about 40--50\textdegree{} and minimum solar activity which represent a relatively conservative (large background) scenario. In this study, the flux given by Equation~\ref{eq:albedomodel} was additionally multiplied by a factor of 1.5, corresponding to the most conservative case of albedo in the polar regions. 

The angular distribution of albedo photons depends on energy, exhibiting central brightening for low photon energies $E<1$\,MeV and limb brightening for $E\geq1$\,MeV~\cite{sazonov2007,mizuno2004}. These effects are not very important for an instrument with a wide field of view and were ignored in this study: the albedo emission as seen by the satellite was assumed to be uniformly distributed over the Earth disk.

The background spectra were simulated within an energy range of 20\,keV--100\,MeV.
Both background sources were defined in the MEGAlib simulations as isotropic emission over the full sphere.
In the analysis stage, the photon directions were constrained to the sky region occupied by Earth for albedo gamma-rays and to the unocculted sky for cosmic gamma rays.
This approach helps to calculate the background count rates for different satellite orientations without rerunning the full Monte Carlo simulations for each particular case.
It does, however, require that the simulated flux is appropriately scaled accounting for the difference between the solid angle of the directional source and the full sphere of the isotropic simulated flux.

\section{Simulation Analysis}
\label{sec:simanalysis}

Each of the three sources (GRB, cosmic background, Earth albedo background) were separately simulated against the EIRSAT-1 geometry using MEGAlib. In addition, an isotropic photon distribution with a flat spectrum having constant intensity between 20\,keV and 3\,MeV was simulated to calculate the GMOD effective area in different energy bands and ensure sufficient photons at higher energies. 

For each simulated photon that has some energy deposited in the sensitive detector volume (scintillator),  MEGAlib records a number of parameters including the event ID, the position, direction and energy of the initial photon, the actual energy that was deposited in the scintillator and the energy that was ``measured'' by the detector taking its energy resolution into account. Two custom file parsers were written to import the useful parameters from MEGAlib's \texttt{.sim} and  \texttt{.tra} output files into a Python Pandas dataframe. The data were then analysed in Python to produce effective area plots, estimate the background count rate and the number of photons detected per unit GRB flux assuming the typical spectrum, and finally to calculate the GRB detection rate based on the distribution of GRBs found in the BATSE 4B catalog~\cite{paciesas1999}.

\subsection{Effective Area}
\label{sec:energyeffectivearea}

\begin{figure}
\includegraphics[width=0.7\textwidth]{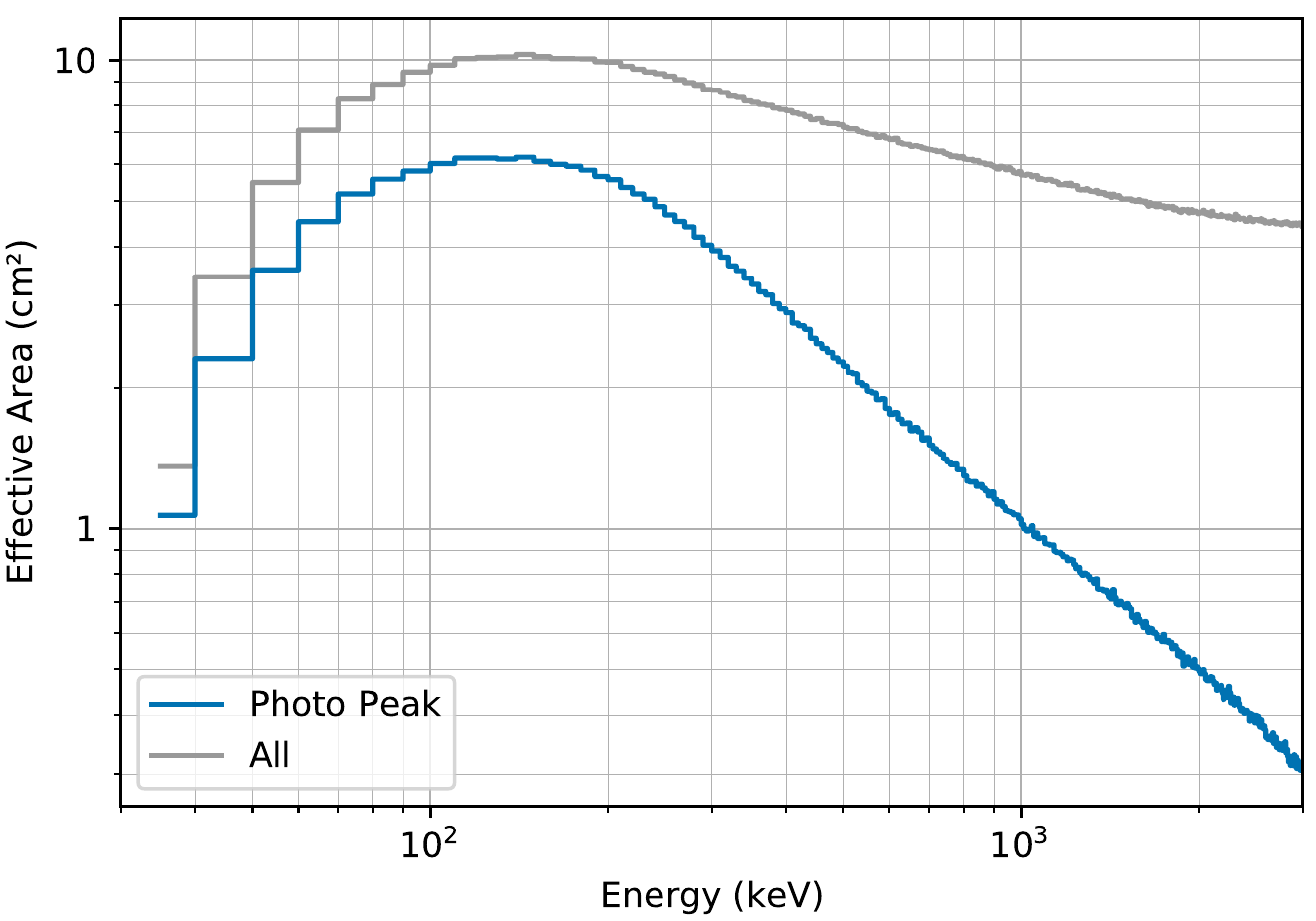}
\caption{The simulated effective area of GMOD in EIRSAT-1, averaged over all directions, as a function of energy. The effective areas were calculated in 10\,keV energy bins. The blue line represents the photo peak efficiency while the grey line includes  photons which were detected at a lower energy.}
\label{fig:effectiveareaenergygraph}
\end{figure}

The effective area of GMOD, averaged over all directions, as a function of energy is shown in Figure~\ref{fig:effectiveareaenergygraph}. The effective area in each energy bin was simulated using a flat photon spectrum.
The blue line gives the photo peak efficiency, while the grey line includes photons which were detected at a lower energy than their initial energy due to scattering or escape effects.
The grey line is a good representation of the effective area for burst detection purposes as the spectral range used for GRB triggering tends to be quite wide and therefore photons recorded at lower than their true energy will still contribute to the count rate in the detector.

The total effective area peaks at approximately 120\,keV reaching a value of 10 cm$^2$. It decreases at lower energies due to the absorption of gamma rays by the spacecraft structure and the aluminium housing of the detector. 
Figure~\ref{fig:effectiveareaenergygraph} indicates that the instrument is most sensitive in the tens of keV up to a few MeV energy range.

\begin{figure}
\includegraphics[width=0.9\textwidth]{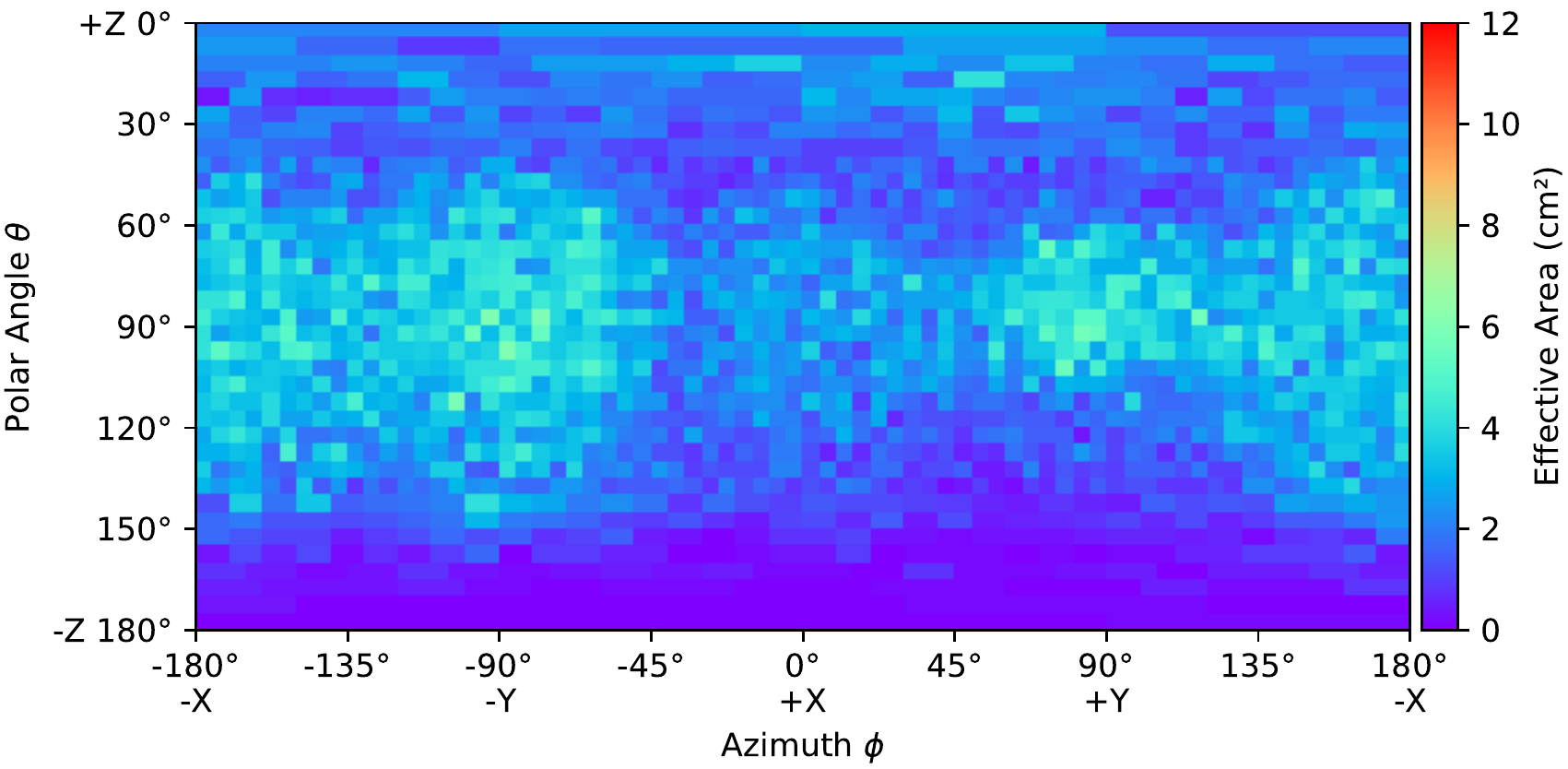}
\begin{center}
(a) 30--50\,keV.
\end{center}

\includegraphics[width=0.9\textwidth]{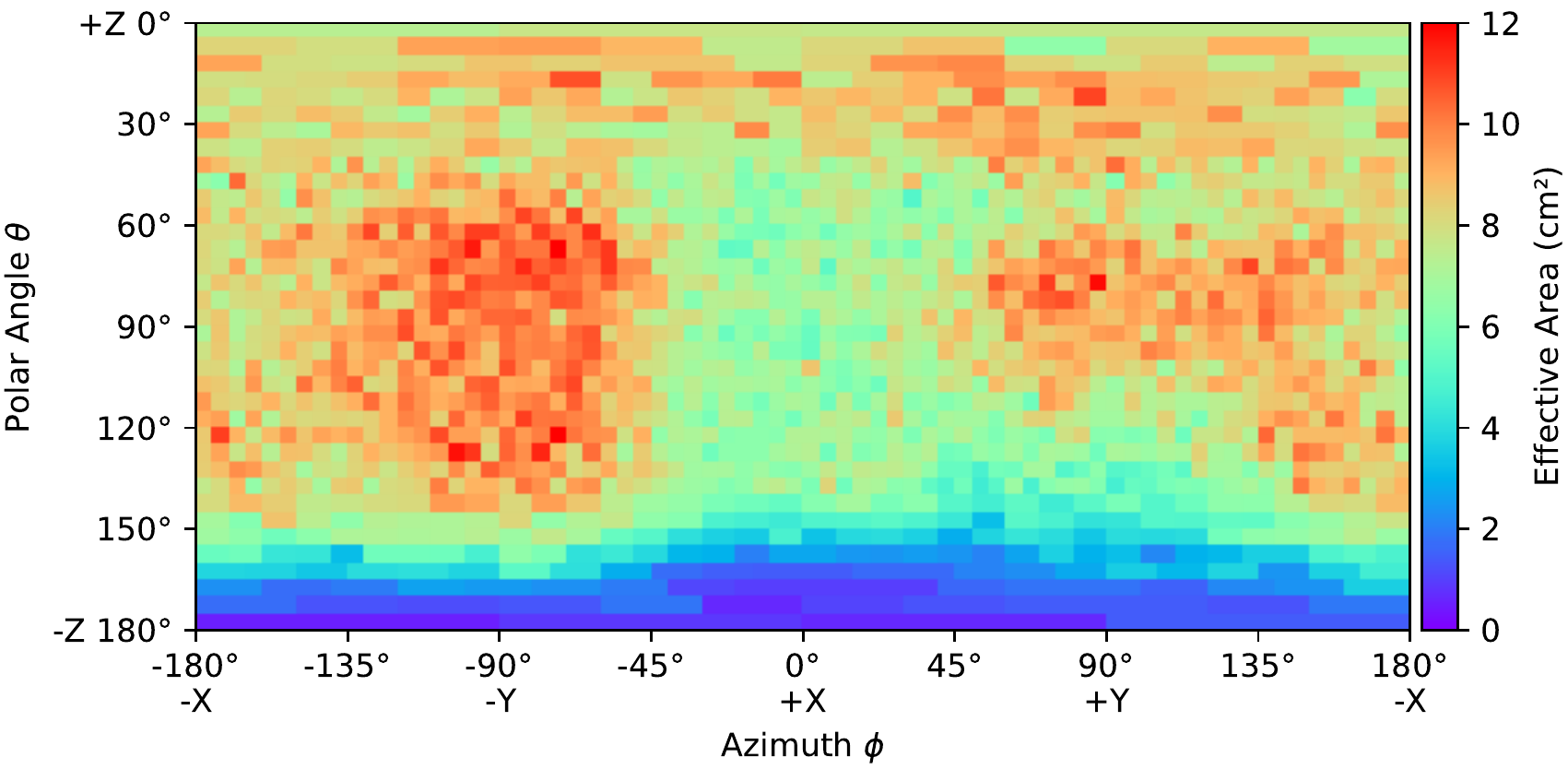}
\begin{center}
(b) 50--100\,keV.
\end{center}

\includegraphics[width=0.9\textwidth]{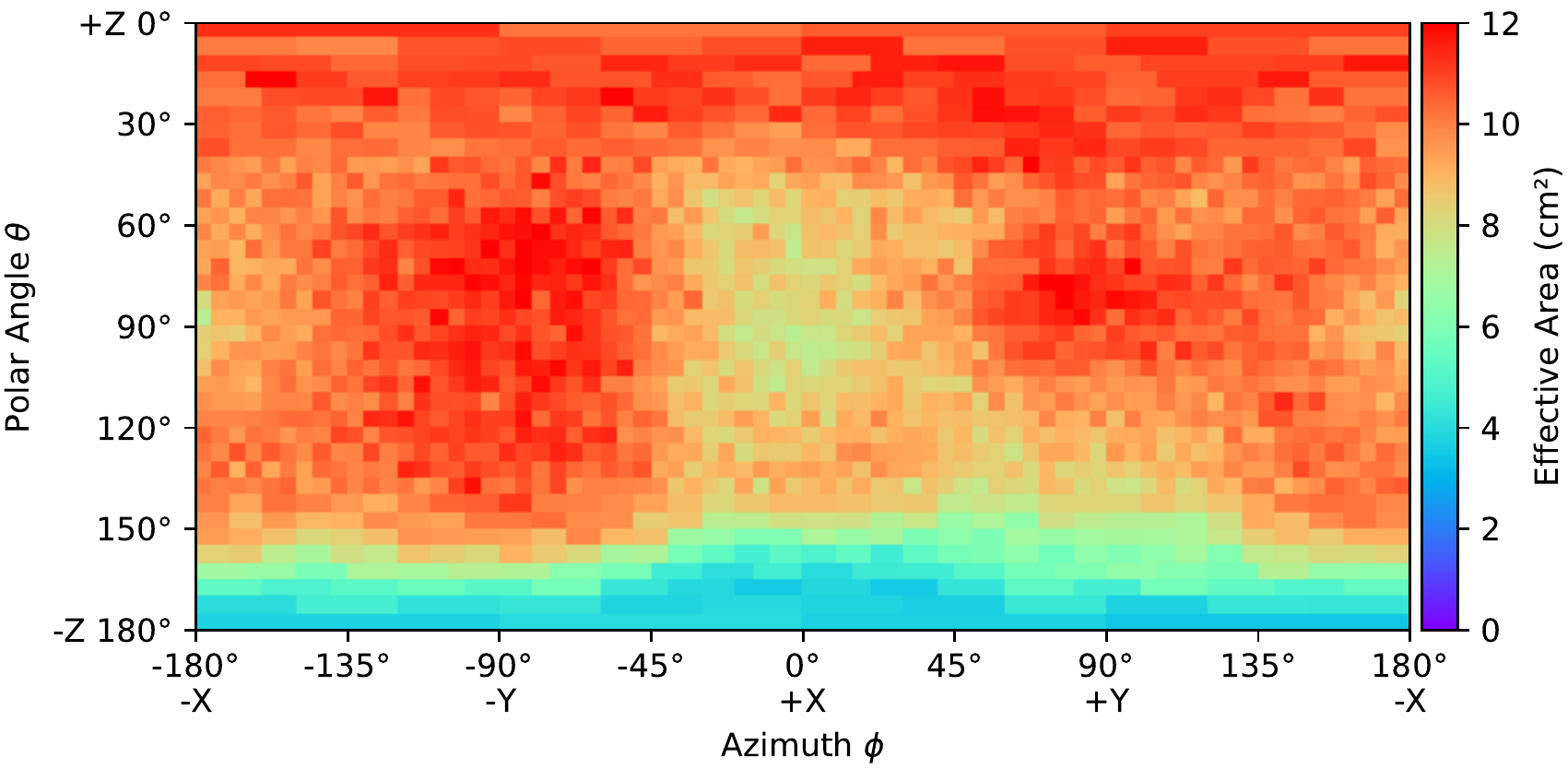}

\begin{center}
(c) 100--300\,keV.
\end{center}

\caption[GMOD Effective Area vs Direction in Various Energy Bands]{The simulated effective area of GMOD in EIRSAT-1 as a function of direction in various energy bands. }
\label{fig:effectiveareaenergybands}
\end{figure}

\begin{figure}
\includegraphics[width=0.9\textwidth]{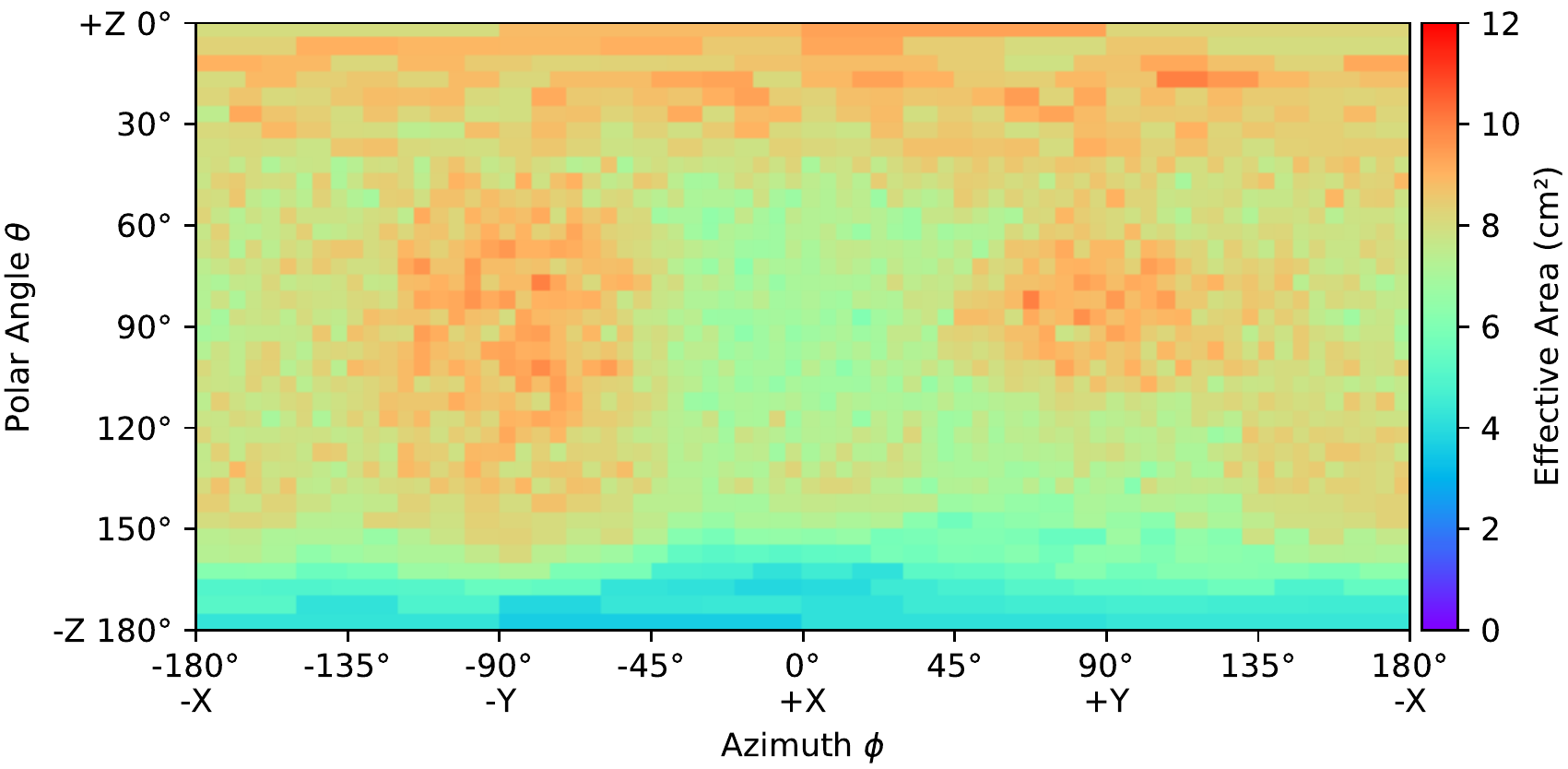}
\begin{center}
(a) 300--500\,keV. 
\end{center}

\includegraphics[width=0.9\textwidth]{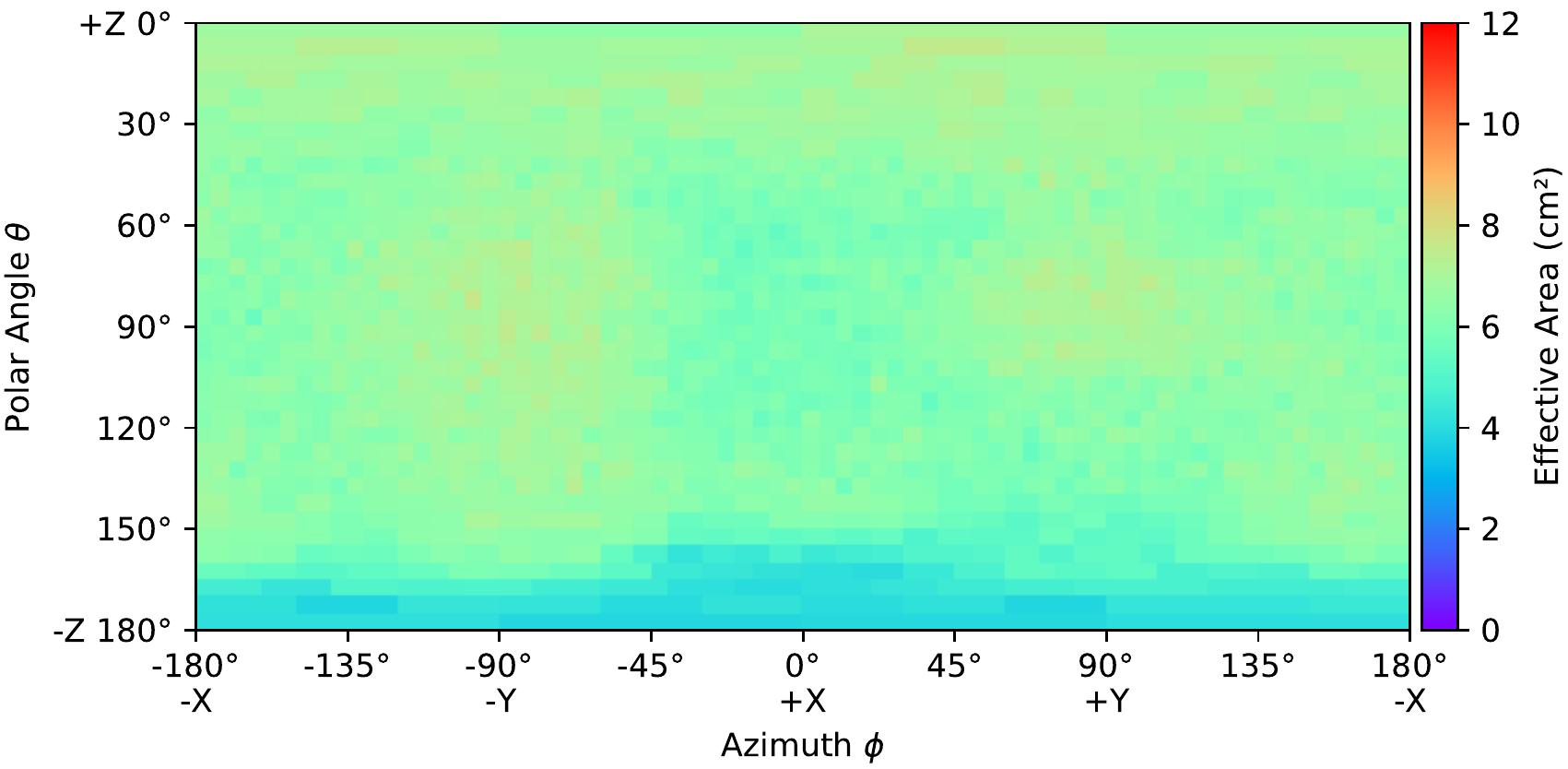}
\begin{center}
(b) 500\,keV--1\,MeV.
\end{center}

\includegraphics[width=0.9\textwidth]{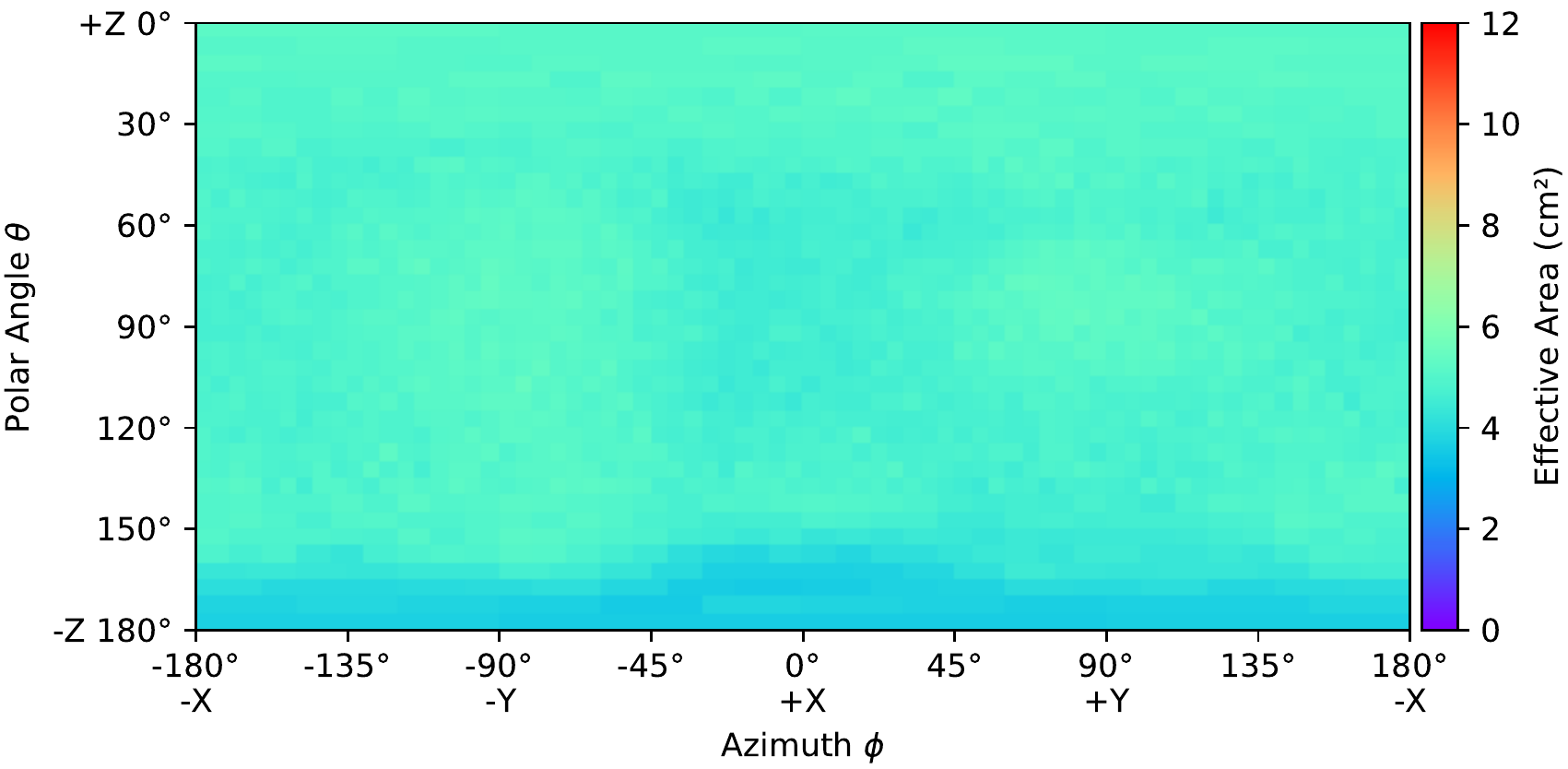}
\begin{center}
(c) 1--3\,MeV.
\end{center}

\caption{The simulated effective area of GMOD in EIRSAT-1 as a function of direction in various energy bands (cont).}
\label{fig:effectiveareaenergybands2}
\end{figure}

The effective area as a function of direction is shown in Figures~\ref{fig:effectiveareaenergybands} and~\ref{fig:effectiveareaenergybands2} for six energy bands. 
The entire field of view is divided into bins (pixels) by azimuthal angle $\phi$ and polar angle $\theta$ and the effective area is calculated for each bin. A 5\textdegree{} bin size is used in both directions. 
This binning suffers from poor simulation statistics at the poles caused by the fact that the solid angles represented by each pixel becomes very small in these regions.
An adaptive rebinning technique was therefore used to increase the size of pixels at the poles such that no pixel had a solid angle more than 1.5 times smaller that those at the equator.
The effect of this rebinning can be seen in the plots below a polar angle of 40\textdegree{} where pixels get progressively wider as they approach 0\textdegree{} and pixels above 140\textdegree{} get progressively wider as they approach 180\textdegree{}. For each pixel, the effective area is the average value calculated using a flat photon spectrum in the given energy band.

Figure~\ref{fig:effectiveareaenergybands} reveals many details about the effect of the structure of the GMOD detector and the EIRSAT-1 spacecraft.
The two most striking features are the high effective area region at $(\theta=90\degree{}$, $\phi=-90\degree{})$, the $-$Y direction (the spacecraft coordinate reference frame is indicated in Figures~\ref{fig:eirsat-1exploded} \& \ref{fig:gmodchopped}), and the very poor effective area at $\theta=180\degree{}$, the $-$Z direction.
The $-$Z direction is blocked by the spacecraft bus and so poor performance in this direction was expected.

The high effective area region in the $-$Y direction corresponds to one of the long faces of the CeBr$_3$ crystal and this direction in particular is where the crystal is closest to edge of the spacecraft.
There is a corresponding smaller region of high effective area in the opposite $+$Y direction.
This region has a cut off at a smaller polar angle than the $-$Y due the effect of the PCB stack attenuating photons.
This imbalance between $\pm$Y would likely be resolved if GMOD were to be placed centrally within the EIRSAT-1 structure.
The geometry explaining this effect can be most clearly seen in the bottom-right part of Figure~\ref{fig:massmodel} which shows a horizontal cross-section through the centre of the crystal.

The $+$Z direction which is shown at $\theta=0\degree{}$ retains reasonably high effective area in the 100--300\,keV band as it also corresponds to one of the long faces of the CeBr$_3$ crystal.
Attenuation through the EMOD thermal coupon assembly (TCA) is however higher than through the solar arrays found in the $\pm$Y directions. This effect is particularly strong in the 30-50\,keV and 50-100\,keV bands and  less obvious in the higher energy bands.
A line of lower effective area can also be seen snaking around the $+$Z face of the spacecraft between the adjacent faces at polar angles between 30\textdegree{} and 60\textdegree{}.
This corresponds to the bulk of aluminium and titanium around the perimeter of the TCA which can be seen in Figures~\ref{fig:eirsat-1exploded} and~\ref{fig:simplifiedcap}.

The $+$X direction at $(\theta=90\degree{}$, $\phi=0\degree{})$ shows a relatively average effective area.
This direction corresponds to one of the smaller, square sides of the crystal and in particular this side is also the one which is coupled to the SiPM array.
This direction therefore exhibits attenuation from both the SiPM array PCB and the ASIC PCB.
There is a matching region in the opposite $-$X direction at $(\theta=90\degree{}$, $\phi=\pm180\degree{})$ direction, also corresponding to a square side of the crystal.
In this case however, there are no SiPMs or PCBs and therefore the effective area remains higher, though not as high as those directions with a long side of the crystal.

Figure~\ref{fig:effectiveareaenergybands2} shows that the effect of the spacecraft structure is smaller at higher energies, as expected.
In the higher energy bands, the photons are energetic enough to penetrate the spacecraft structure from almost any angle equally, though the effective area is reduced as the scintillator absorption is also lower.

%
\subsection{GRB Effective Area}
\label{sec:grbeffectivearea}
Similar to BATSE and GBM, GMOD will detect GRBs by registering a statistically significant increase in the detector count rate. 
For the purposes of burst detection in this study, the GMOD count rate is integrated over an energy range of 50--300\,keV. This is the range of maximum effective area for GMOD (Figure~\ref{fig:effectiveareaenergygraph}) comprising the majority of detected photons from a typical GRB. It was also the energy range of the nominal BATSE on-board burst trigger used during the first several years of BATSE operation~\cite{paciesas1999}. The GRB peak fluxes in the BATSE catalog are given in this range. Calculation of the GMOD count rate for these GRBs is less sensitive to their spectra if performed in the identical energy band, 50--300\,keV. 

The GRB effective area, $A_{\textrm{GRB}}$, is defined here as the total number of photons detected with a reconstructed energy in the 50--300\,keV energy band, $N_{\textrm{det, band}}$, divided by the total number of photons simulated per cm$^2$ with an initial energy in that same energy band, $H_{\textrm{band}}$. 
\begin{equation}
  A_{\textrm{GRB}} = \frac{N_{\textrm{det, band}}}{H_{\textrm{band}}}
\label{eq:effareasimple}
\end{equation}
In other words, this quantity represents the detector count rate in the 50--300\,keV energy band per unit GRB flux in the 50--300\,keV band. Unlike the effective area described in Section~\ref{sec:energyeffectivearea} (which was simply averaged over photons simulated in the given energy bands), this quantity includes contributions to the 50--300\,keV count rate from the entire GRB spectrum. 

The GRB effective area of GMOD was calculated for the typical GRB spectrum described in Section~\ref{sec:grbmodel}. The simulated fluence in the 50--300\,keV band can be calculated from the total simulated fluence $H_{\textrm{tot}}$ as 
\begin{equation}
  H_{\textrm{band}} = R_{\textrm{band}} H_{\textrm{tot}},
\label{eq:simfluence}
\end{equation}
where $R_{\textrm{band}}$=0.493 is the ratio of photons in the simulated source spectrum that have an energy in the 50--300\,keV band. 

Therefore the effective area can be rewritten in terms of known parameters from the simulation as
\begin{equation}
  A_{\textrm{GRB}} = \frac{ N_{\textrm{det, band}} }{ R_{\textrm{band}} H_{\textrm{tot}} }
\label{eq:effareaknown}
\end{equation}

\begin{figure}
\includegraphics[width=0.9\textwidth]{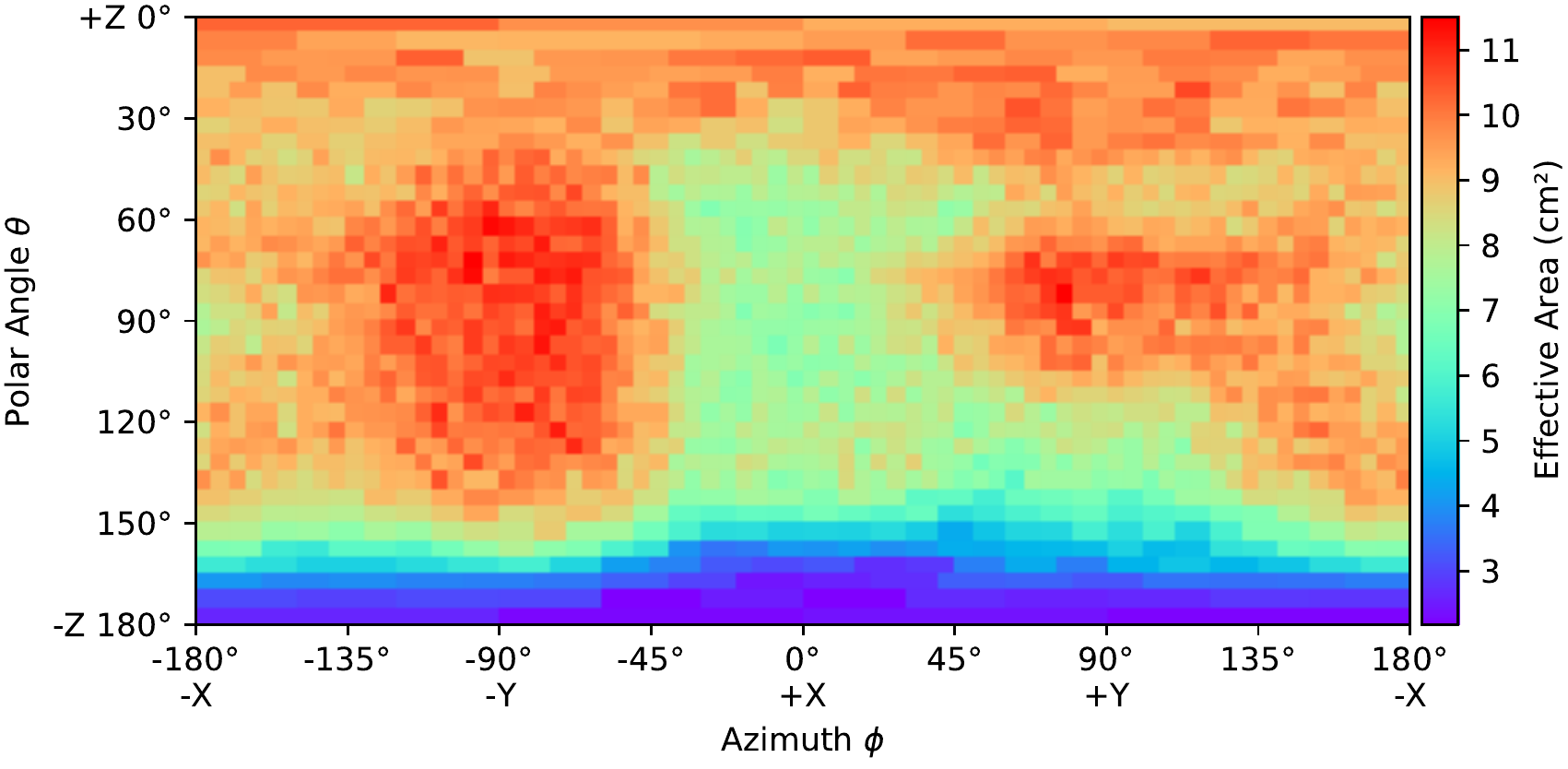}
\caption[GMOD Effective Area vs Direction]{The GRB effective area of GMOD in EIRSAT-1 as a function of direction. The effective area is calculated in the 50--300\,keV range for a typical GRB spectrum ($\alpha=-1.1$, $\beta=-2.3$, $E_\mathrm{peak}=300$\,keV).}
\label{fig:effectivearea}
\end{figure}

The GRB effective area of GMOD on EIRSAT-1 as a function of azimuth and polar angle is shown in Figure~\ref{fig:effectivearea}.
As expected, the plot displays similar structural effects as were seen in  Figure~\ref{fig:effectiveareaenergybands} for the effective area in the 50--100\,keV and 100--300\,keV energy bands. The average value of the GRB effective area calculated over all directions is 8.75\,cm$^2$.

\subsection{Effects of Spacecraft Spin and Earth Occultation}
\label{sec:spin}

As the Earth occults a large part of the sky (approximately 33\% at 400\,km altitude), the average effective area of GMOD over the unnocculted sky depends on satellite orientation with respect to the planet. The orientation also affects the detector count rate from cosmic and albedo gamma rays. 

Due to the constraints of the magnetorquer based attitude control system on EIRSAT-1, the spacecraft will be spin stabilised at a few to tens of revolutions per minute. The effective area and background, averaged over the rotation period, depend only on the angle between the +Z spin-axis of the spacecraft and Zenith.  

The effect of the spin of EIRSAT-1 can be accounted for by aligning the azimuth axis of the spherical coordinates used in the simulation with the +Z spin-axis, averaging over all azimuthal angles and applying a weight based on the fraction of time that any given polar angle relative to the spacecraft is oriented toward or away from Earth.
In the cases of Earth-occulted sources, such as GRBs or cosmic background, the weight will be the fraction of the revolution that a given polar angle relative to the spacecraft spends oriented away from the Earth, while for albedo background it will be the fraction spent pointed towards the Earth.

In order to calculate these weights, the Earth is modelled as a disc of a given angular size.
The extents of this disc form a circle projected onto a sphere in the spherical coordinate reference frame of the spacecraft.
The centre of the disc is the spacecraft-Earth vector and its radius, being the shortest distance between the centre and the edge of the disc, is therefore a great circle on the sphere.

\begin{figure}
\begin{center}
  \includegraphics[width=0.5\textwidth]{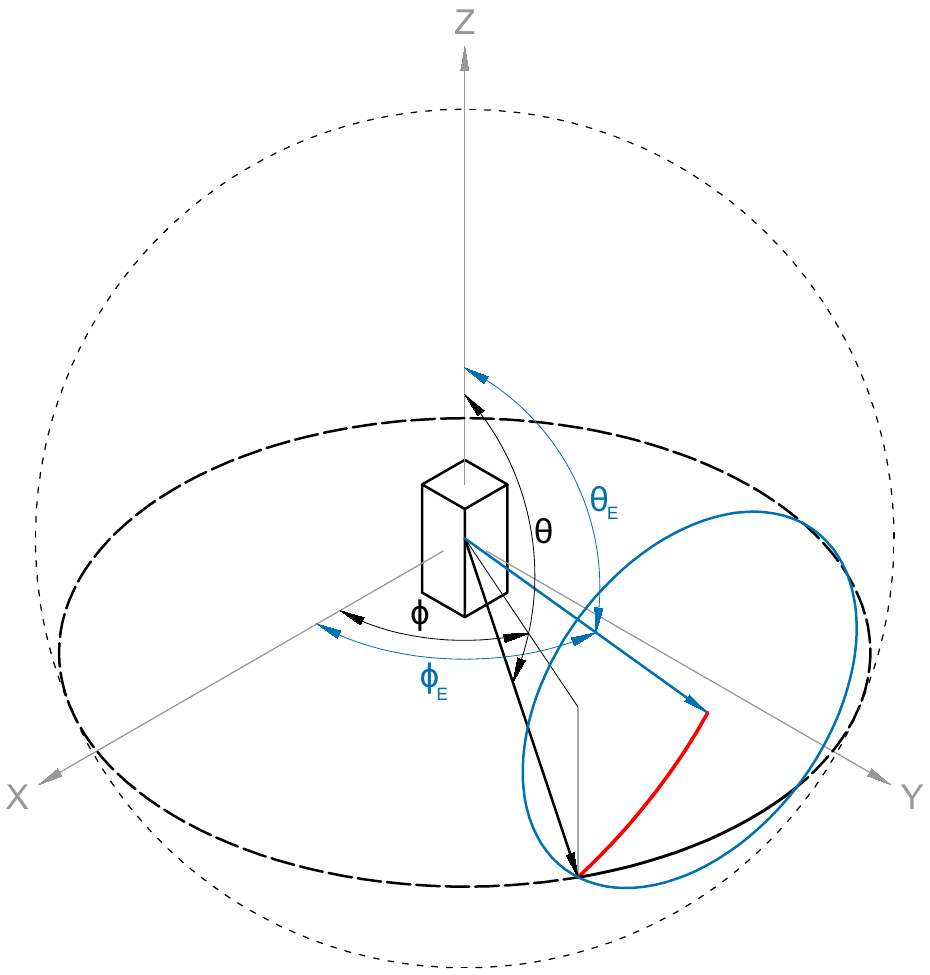}
  \caption{Geometry used to calculate the fraction of time any direction from the spacecraft is oriented towards the Earth. The blue vector ($\theta_E$, $\phi_E$) points towards Earth, with the blue circle giving the maximum extents of the Earth and its angular radius in red. For a vector ($\theta_E$, $\phi_E$) such as that shown in black, the solid black line inside the blue circle represents the fraction of the full revolution spent pointed towards Earth. Note: for clarity, the Earth disc is shown as a smaller angular size than reality.}
  \label{fig:spin}
\end{center}
\end{figure}

This geometry is shown in Figure~\ref{fig:spin}.
The blue vector is the spacecraft-Earth vector ($\theta_E$, $\phi_E$), with the blue circle indicating the extents of the apparent disc of the Earth and the red arc indicating the great circle distance radius of the Earth.
The black vector is an orientation in the spacecraft frame from which a photon may originate ($\theta$, $\phi$), with the thicker dashed black circle indicating the arc which it sweeps out across the sky.
The portion of the revolution that that vector spends oriented towards the Earth is the solid black arc inside the blue circle.
This angle swept out in this portion is twice the difference between the azimuthal angles, $\Delta\phi$.
The weight, $W$, corresponding to the fraction of the revolution spent pointed towards the Earth is therefore
\begin{equation}
  W = \frac{2\left(\phi_E - \phi\right)}{2\pi} = \frac{\Delta\phi}{\pi}
\label{eq:weight}
\end{equation}

The great circle distance formula gives the shortest distance between two points on a sphere.
For a unit sphere, this distance is equivalent to the angle subtended on the sphere between two points, or in this problem the angular radius of the Earth as seen from the spacecraft.

The great circle distance, $\psi$, between two points ($\theta_E,\phi_E$) and ($\theta,\phi$) in spherical coordinates is given by:
\begin{equation}
  \psi = \cos^{-1} \left( \cos\theta_E \cos\theta + \sin\theta_E \sin\theta\cos\left( \phi_E - \phi \right) \right)
\label{eq:greatcircle}
\end{equation}

Equation~\ref{eq:greatcircle} is thus rearranged to give:
\begin{equation}
  \Delta\phi = \cos^{-1} \left( \frac{\cos\psi - \cos\theta_E \cos\theta}{\sin\theta_E \sin\theta} \right),
\label{eq:phifrac}
\end{equation}
which is only valid when $|\theta - \theta_E| \leq \psi/2$, representing parts of the spacecraft which see the Earth for only part of the revolution.
Where the Earth is seen for the full revolution, the bracketed term in Equation~\ref{eq:phifrac} evaluates as $<-1$ and where the Earth is not seen at all, the bracketed term evaluates as $>1$, both of which are undefined for real values for the inverse cosine function.
Clipping the range of the bracketed term to between -1 and 1 gives the appropriate $\Delta\phi$ for the fully Earth or fully space orientations.
Finally, dividing $\Delta\phi$ by $\pi$ as per Equation~\ref{eq:weight} gives $W$, the fraction of a revolution that any part of the spacecraft, with a polar angle of $\theta$ in the spacecraft's spherical coordinate reference frame, spends pointed towards the Earth.

\subsection{GRB Effective Area as a Function of Spacecraft Attitude}
\label{sec:attitudeeffectivearea}

Utilising the technique discussed in Section~\ref{sec:spin}, it is possible to simulate the effect of Earth occultation on the average effective area of GMOD.
In this analysis the Earth is placed at a range of polar angles in the spacecraft's reference frame, $180\degree{} \geq \theta_E \geq 0\degree{}$, representing pointing angles 0\degree{}--180\degree{} between the $+$Z axis and Zenith.
For an altitude of approximately 400\,km above the Earth, the Earth appears as a disc which subtends an angular radius of 70\textdegree{} and occults 33\% of the sky.

The GRB effective area is calculated over all directions as before but rather than simply counting the detected photons, each photon carries a weight and it is the weights of detected photons that are summed.
The photon weights are the fraction of a revolution for which the direction in the spacecraft's rotating frame from which they originated, was pointed towards space. The total simulated fluence $H_{\textrm{tot}}$ in Equation~\ref{eq:effareaknown} is replaced by $0.67H_{\textrm{tot}}$ representing the total fluence of unocculted photons.

\begin{figure}
\includegraphics[width=0.8\textwidth]{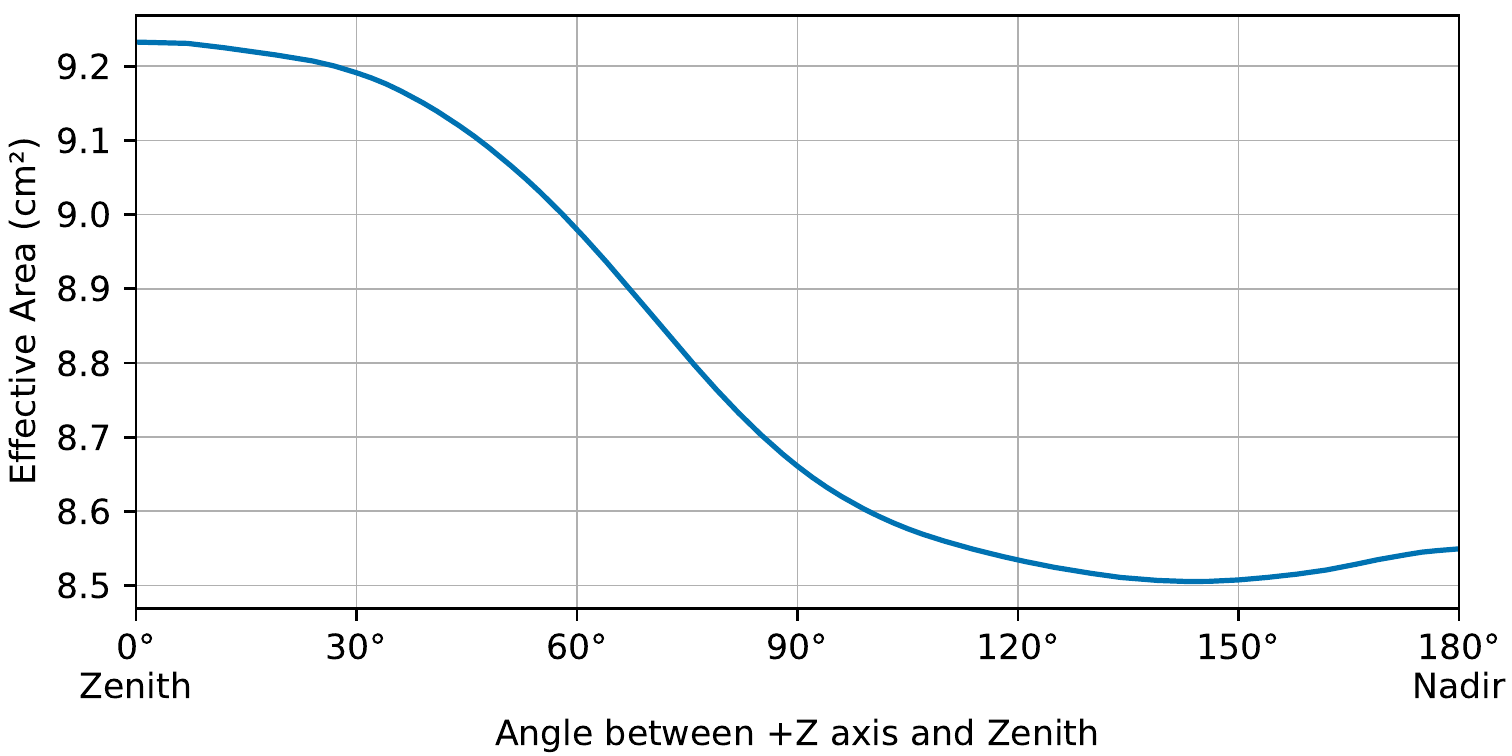}
\caption[GMOD Effective Area vs Spacecraft Attitude]{The GRB effective area of GMOD in EIRSAT-1, averaged over the entire unocculted sky, as a function of spacecraft attitude. The effective area is calculated in the 50--300\,keV range for a typical GRB spectrum.}
\label{fig:effectiveareaattitude}
\end{figure}

The resulting average effective area of the instrument as a function of spacecraft attitude is shown in Figure~\ref{fig:effectiveareaattitude}. As expected, the average effective area is maximised when the least sensitive ($-$Z) side of the detector is facing the Earth. However, this effect is relatively small. 

\subsection{Background Rate}
\label{sec:bgrate}
\begin{figure}
  \includegraphics[width=0.8\textwidth]{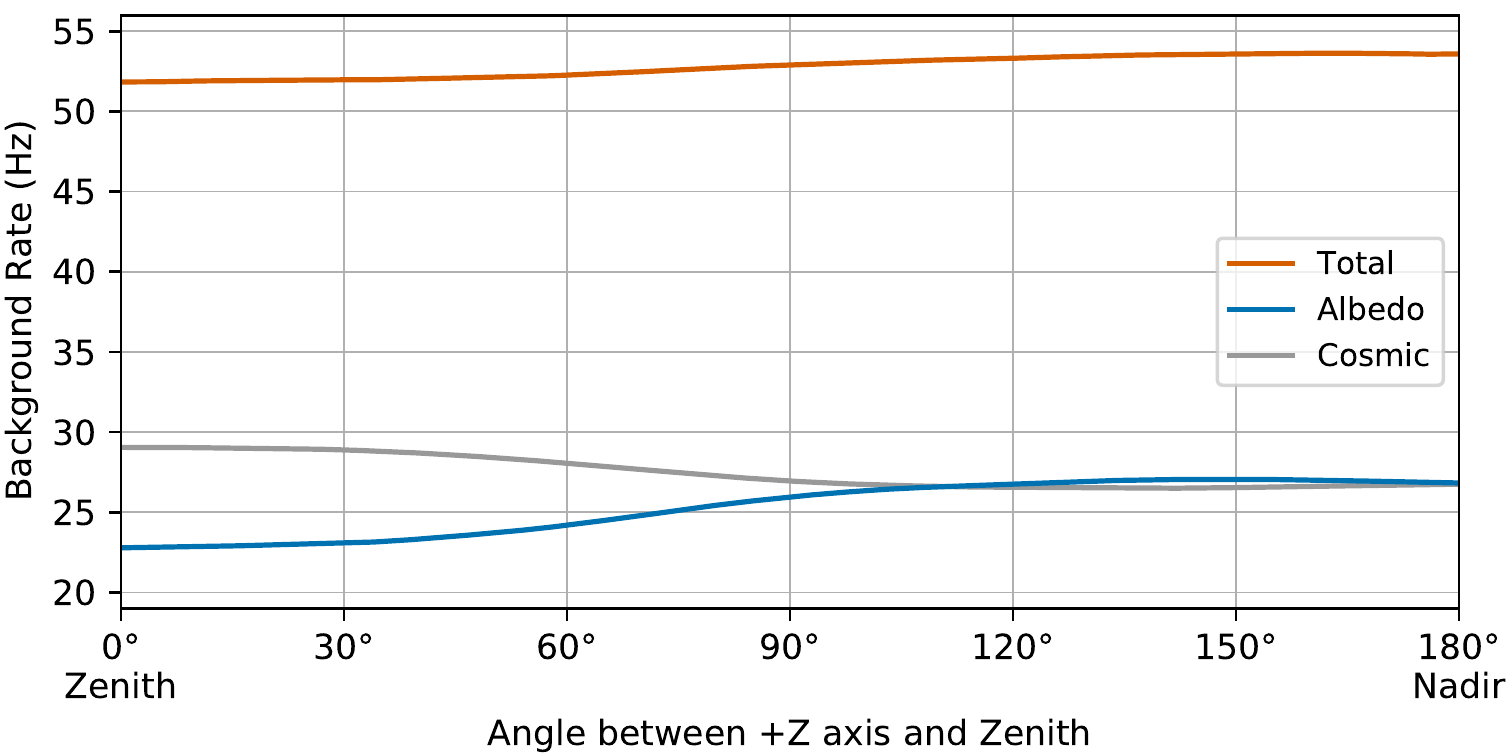}
  \caption[GMOD Background Count Rates vs Spacecraft Attitude]{The simulated background count rates in GMOD in the 50--300\,keV range with contributions from cosmic gamma-ray and Earth albedo backgrounds, as a function of spacecraft attitude.}
  \label{fig:attitudenoise}
\end{figure}

The GMOD background count rate was simulated using the cosmic gamma-ray background and Earth albedo sources as described in Section~\ref{sec:bgmodel}.
The background rate averaged over the spacecraft rotation period was calculated for spacecraft pointing angles 0\textdegree{}--180\textdegree{} between the $+$Z axis and Zenith using the technique described in Section~\ref{sec:spin}.
The detected cosmic background photons were weighted by the fraction of a spacecraft revolution spent pointed towards space while the Earth albedo photons were weighted by the fraction spent pointed towards Earth.
The resulting count rates are shown in Figure~\ref{fig:attitudenoise}.

%

\subsection{GRB Detection}
\label{sec:grbdetection}

GMOD's ability to detect GRBs was analysed by calculating the detection significance which would be expected for GRBs with the same fluxes as those found in the BATSE 4B catalog~\cite{paciesas1999}. 

The number of photons which would be detected by GMOD from a given GRB depends on the GRB position in the spacecraft coordinate frame. To simplify the calculations, 
the number of detected counts was determined using the average GRB effective area of the detector as a function of spacecraft attitude as described in Section~\ref{sec:attitudeeffectivearea}.
For each GRB in the BATSE 4B catalog, the flux for the 64\,ms, 256\,ms, and 1024\,ms trigger bins was multiplied by the effective area and then scaled by the bin duration to get the number of counts detected from the GRB.
The GMOD background count rates were also scaled by the bin duration.
The detection significance was then calculated as the GRB counts divided by the square root of the background counts.
The detection significance, $\sigma$ is 
\begin{equation}
  \sigma = \frac{ A_{\textrm{GRB}} F t_\textrm{bin} }{ \sqrt{B t_\textrm{bin} } },
\label{eq:significance}
\end{equation}
where $A_{\textrm{GRB}}$ is the GMOD effective area for GRBs, $F$ is the flux reported by BATSE, $t_\textrm{bin}$ is the duration of the trigger bin, and $B$ is the background count rate for GMOD. Using the effective area simulated with the average GRB spectrum, this approach ignores variations in GRB spectral properties.
The significance in each of the three trigger bins as well as the best calculated significance for each GRB was recorded.

The BATSE 4B catalog covers a duration of 1960 days or 5.3 years.
The average exposure factor for the 4B catalog, accounting for Earth occultation and instrument down time due to telemetry gaps or South Atlantic Anomaly (SAA) passage for example, is 0.483.
Therefore if BATSE was able to view the entire sky at once without the effects of Earth occultation and was operational continuously, it would have detected the 1637 GRBs found in the 4B catalog in 2.56 years.

The detection significances are binned using a cumulative histogram with each bin containing the number of GRBs with a significance greater than the lower bin edge value.
The values in each bin are then divided by the duration of the catalog, divided by the average BATSE exposure factor and multiplied by the GMOD exposure factor of 0.496 to give the GMOD GRB detection rate as a function of significance.
This GMOD exposure factor accounts for Earth occultation of $\sim$33\% and the effects of instrument/trigger down-time which will average $\sim$10\% due to transits of the SAA and $\sim$16\% due to transits of the outer Van Allen belt at high latitudes.

\begin{figure}
\includegraphics[width=0.8\textwidth]{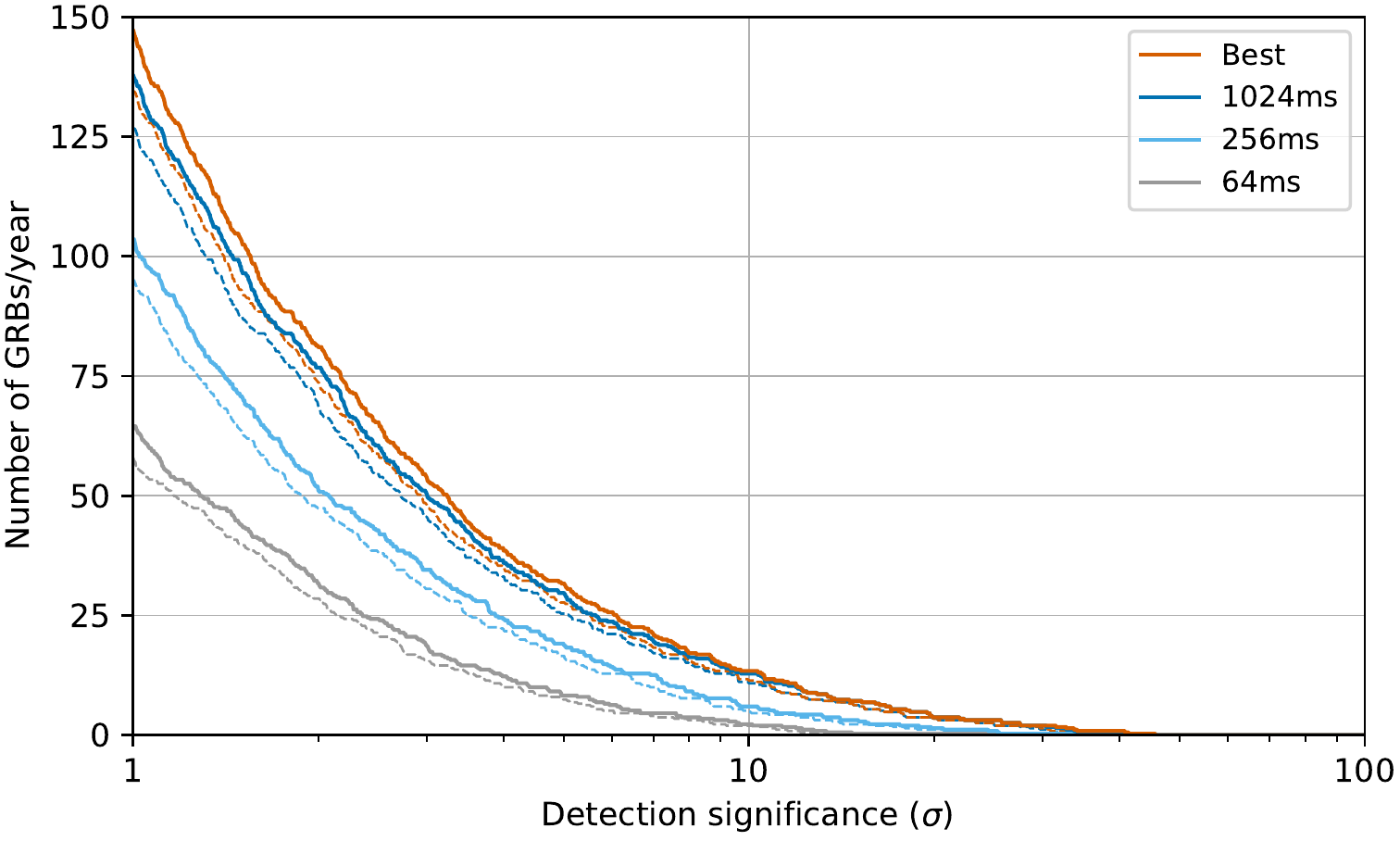}
\caption[GMOD GRB Detection Rate]{The GRB detection rate for GMOD as a function of detection significance for the one-year nominal mission lifetime. The detection rates based on three different trigger bin durations is shown as well as when the best response across the three bins is chosen. The solid lines indicate the detection rate if the spacecraft is always zenith pointing while the dashed lines indicate a pointing angle of 148\textdegree{} away from zenith which is where GMOD's response is most impacted by Earth occultation.}
\label{fig:grbdetection}
\end{figure}

The GMOD GRB detection rates for the nominal mission lifetime of one year are shown in Figure~\ref{fig:grbdetection}.
At a detection significance of 10$\sigma$ GMOD is expected to detect between 11 and 14 GRBs per year, depending of spacecraft attitude.
Many more GRBs will be detected at lower significance.
At 5$\sigma$, the detection rate would be between 28 and 32 GRBs per year.

The detection rates for the shorter trigger bins are noticeably lower, with the 1024\,ms bin very closely matching the detection rate when the best response from all bins is chosen.
This is a particularly interesting result as it indicates that in the event that the on-board processing resources are too constrained to support multiple bins for triggering, the longer 1024\,ms bin is capable of providing a trigger for most GRBs that would be detected using the shorter bins. It should be noted, however, that triggering with the 1024\,ms bin would be inefficient for very short bursts. 

\section{Summary}

GMOD is a small gamma-ray detector designed for a 2U CubeSat. The main purpose of the instrument is to qualify the new detector technology using a CeBr$_3$ scintillator, silicon photomultipliers and the SIPHRA readout ASIC for space applications and to validate its capability to detect GRBs in low Earth orbit.

The sky-average effective area of GMOD, reaching a peak value of 10\,cm$^2$ at 120\,keV, is an order of magnitude lower than that of the GBM NaI detectors or detectors in recently proposed larger GRB-detecting CubeSat missions such as HERMES, CAMELOT, MoonBeam, GRID and BurstCube. The shape of the scintillator in GMOD results in omni-directional sensitivity which allows for a nearly all-sky field of view but also limits the GRB sensitivity in any given direction. Thin scintillators used in other missions provide better sensitivity for directions close to the detector normal. The limited field of view of such detectors can be compensated by using several detectors oriented in different directions on board a single spacecraft and/or using a network of satellites viewing different parts of the sky.

Despite the relatively small effective area, GMOD is expected to detect between 11 and 14 GRBs, at a significance greater than 10$\sigma$ (and up to 32 at 5$\sigma$), during a nominal one-year mission.
It will be able to record the light curves of the brightest bursts and measure their spectra in an energy range from tens of keV up to about 1\,MeV. This will be an important step in technology qualification which will prepare the ground for design of larger detectors and future instruments.

\begin{acknowledgements}
The EIRSAT-1 project is carried out with the support of ESA’s Education Office under the Fly Your Satellite!\,2 programme. This study was supported by The European Space Agency's Science Programme under contract 4000104771/11/NL/CBi. JM, AU, DM, and SMB acknowledge support from Science Foundation Ireland (SFI) under grant number 17/CDA/4723. LH acknowledges support from SFI under grant 19/FFP/6777 and the EU AHEAD2020 project (grant agreement 871158). DM, RD, MD and JT acknowledge support from the Irish Research Council (IRC) under grants GOIPG/2014/453, GOIPG/2019/2033, GOIPG/2018/2564 and GOIPG/2014/685. We acknowledge all students who have contributed to EIRSAT-1.
\end{acknowledgements}

\bibliographystyle{spphys}       
\bibliography{refs.bib}   

\end{document}